\documentclass[11pt]{article}
\pdfoutput=1

\usepackage{cancel,slashed}
\usepackage{bbm}
\usepackage{amssymb}
\usepackage{amsfonts}
\usepackage{amsmath}
\usepackage{graphicx}
\usepackage{cite}
\usepackage{physics}

\usepackage{ dsfont }
\usepackage{latexsym}
\usepackage{color}
\usepackage{cancel,slashed}
\usepackage[hyperfootnotes=false,linktocpage]{hyperref}
\usepackage{color}
\usepackage{transparent}
\usepackage[hang,flushmargin]{footmisc}

\usepackage{blkarray}
\usepackage{multirow}
\usepackage{empheq}

\usepackage{comment}
\usepackage{float}

\newcommand{\bmat}{\left(\begin{array}}
\newcommand{\emat}{\end{array}\right)}

\newcommand{\pr}{\mathbbm{R}}
\newcommand{\pz}{\mathbbm{Z}}

\def\CK {{\cal K}}
\def\a {\alpha}

\def\1{{\bf 1}}
\def\2{{\bf 2}}
\def\3{{\bf 3}}
\def\4{{\bf 4}}
\def\6{{\bf 6}}

\def\targ#1#2{\genfrac{[}{]}{0pt}{}{#1}{#2}}
\def\targ2#1#2{\genfrac{}{}{0pt}{}{#1}{#2}}

\definecolor{mygr}{rgb}{0,0.6,0}
\definecolor{mygrey}{rgb}{0,0.1,0.2}
\definecolor{myblue}{rgb}{0,0.5,0.9}
\definecolor{myblue2}{rgb}{0,0.5,0.5}
\definecolor{myblue3}{rgb}{0,0.7,0.9}
\definecolor{myblue4}{rgb}{0,0.6,0.6}
\definecolor{myorange}{rgb}{1,0.5,0}
\definecolor{mypurple}{rgb}{0.6,0,1}
\definecolor{mygolden}{rgb}{1,0.8,0.2}
\definecolor{mycyan}{rgb}{0,1,1}
\definecolor{mymagenta}{rgb}{1,0,1}
\definecolor{mykiwi}{rgb}{0.8,1,0.5}
\definecolor{mybrown}{cmyk}{0.14, 0.42, 0.56, 0.2}
\definecolor{myturq}{cmyk}{0.99, 0, 0.2, 0.4}
\definecolor{myaubergine2}{cmyk}{0.4, 0.5, 0, 0.1}
\definecolor{myaubergine}{cmyk}{0.6,0.85,0,0}
\definecolor{CycleGreen}{cmyk}{0.52,0,1,0}
\definecolor{CycleBrown}{cmyk}{0, 0.4, 0.9, 0.2}

\DeclareFontFamily{U}{rcjhbltx}{}
\DeclareFontShape{U}{rcjhbltx}{m}{n}{<->rcjhbltx}{}
\DeclareSymbolFont{hebrewletters}{U}{rcjhbltx}{m}{n}

\DeclareMathSymbol{\lamed}{\mathord}{hebrewletters}{108}
\DeclareMathSymbol{\mem}{\mathord}{hebrewletters}{109}
\DeclareMathSymbol{\ayin}{\mathord}{hebrewletters}{96}
\DeclareMathSymbol{\tsadi}{\mathord}{hebrewletters}{118}
\DeclareMathSymbol{\qof}{\mathord}{hebrewletters}{113}
\DeclareMathSymbol{\resh}{\mathord}{hebrewletters}{114}
\DeclareMathSymbol{\pe}{\mathord}{hebrewletters}{112}
\DeclareMathSymbol{\pesofit}{\mathord}{hebrewletters}{80}
\DeclareMathSymbol{\samekh}{\mathord}{hebrewletters}{115}
\DeclareMathSymbol{\tav}{\mathord}{hebrewletters}{116}
\DeclareMathSymbol{\vav}{\mathord}{hebrewletters}{119}
\DeclareMathSymbol{\het}{\mathord}{hebrewletters}{120}
\DeclareMathSymbol{\yod}{\mathord}{hebrewletters}{121}
\DeclareMathSymbol{\zayin}{\mathord}{hebrewletters}{122}
\DeclareMathSymbol{\alephdot}{\mathord}{hebrewletters}{128}
\DeclareMathSymbol{\tsadisofit}{\mathord}{hebrewletters}{90}
\DeclareMathSymbol{\shin}{\mathord}{hebrewletters}{152}

\newcommand{\bk}[1]{{\color{black}  #1}}

\def\CN {{\cal N}}

\def\d{{\delta}}
\def\be{\begin{equation}}
\def\ee{\end{equation}}
\def\bea{\begin{eqnarray}}
\def\eea{\end{eqnarray}}
\def\bes{\begin{subequations}}
\def\ees{\end{subequations}}

\def\eps{{\epsilon}}
\def\oh{\frac{1}{2}}

\def\re{\mbox{Re}\, }
\def\im{\mbox{Im}\, }

\def\IZ{\mathbb{Z}}

\def\Om{\Omega}
\usepackage{multicol}
\usepackage{float}
\usepackage{caption}
\def\p {{\partial}}

\def\g {{\gamma}}

\newcommand{\cF}{\mathcal{F}}

\newcommand{\cK}{\mathcal{K}}

\newcommand{\cN}{\mathcal{N}}
\newcommand{\cO}{\mathcal{O}}

\newenvironment{eqn*}{\begin{equation*}\begin{aligned}}{\end{aligned}\end{equation*}\noindent}

\makeatletter
\newsavebox\myboxA
\newsavebox\myboxB
\newlength\mylenA

\newcommand*\xoverline[2][0.75]{%
\sbox{\myboxA}{$\m@th#2$}%
\setbox\myboxB\null
\ht\myboxB=\ht\myboxA%
\dp\myboxB=\dp\myboxA%
\wd\myboxB=#1\wd\myboxA
\sbox\myboxB{$\m@th\overline{\copy\myboxB}$}
\setlength\mylenA{\the\wd\myboxA}
\addtolength\mylenA{-\the\wd\myboxB}%
\ifdim\wd\myboxB<\wd\myboxA%
   \rlap{\hskip 0.5\mylenA\usebox\myboxB}{\usebox\myboxA}%
\else
    \hskip -0.5\mylenA\rlap{\usebox\myboxA}{\hskip 0.5\mylenA\usebox\myboxB}%
\fi}
\makeatother

\begin{document}
\pagestyle{plain}

\makeatletter
\@addtoreset{equation}{section}
\makeatother
\renewcommand{\theequation}{\thesection.\arabic{equation}}

\pagestyle{empty}
\rightline{IFT-UAM/CSIC-22-77}
\vspace{0.5cm}
\begin{center}
\Huge{{New instabilities for  non-supersymmetric \\ AdS$_4$ orientifold vacua} 
\\[10mm]}
\normalsize{Fernando Marchesano,$^1$ Joan Quirant,$^1$ and Matteo Zatti$^{1,2}$ \\[12mm]}
\small{
${}^1$Instituto de F\'{\i}sica Te\'orica UAM-CSIC, c/ Nicol\'as Cabrera 13-15, 28049 Madrid, Spain \\[2mm] 
${}^2$Departamento de F\'{\i}sica Te\'orica, 
Universidad Aut\'onoma de Madrid, 
28049 Madrid, Spain
\\[8mm]} 
\small{\bf Abstract} \\[5mm]
\end{center}
\begin{center}
\begin{minipage}[h]{15.0cm}

We consider massive type IIA orientifold compactifications of the form AdS$_4 \times X_6$, where $X_6$ admits a Calabi--Yau metric and is threaded by background fluxes. From a 4d viewpoint,  fluxes generate a  potential whose vacua have been classified, including one $\cN=1$ and three perturbatively stable $\cN=0$ branches. We reproduce this result from a 10d viewpoint, by solving the type IIA equations at the same level of \bk{detail} as previously done for the $\cN=1$ branch. All solutions exhibit localised sources and parametric scale separation.  We then analyse the non-perturbative stability of the $\cN=0$ branches. We consider new 4d membranes, obtained from wrapping D8-branes on $X_6$ or D6-branes on its divisors, threaded by non-diluted worldvolume fluxes. Using them we show that all branches are compatible with the Weak Gravity Conjecture for membranes. In fact, most vacua satisfy the sharpened  conjecture that predicts superextremal membranes in $\cN=0$ settings, except for a subset whose non-perturbative stability remains an open problem.

\end{minipage}
\end{center}
\newpage
\setcounter{page}{1}
\pagestyle{plain}
\renewcommand{\thefootnote}{\arabic{footnote}}
\setcounter{footnote}{0}


\tableofcontents

\section{Introduction}
\label{s:intro}

AdS vacua are a key sector of the string Landscape. On the one hand, stable vacua should have a dual holographic description that allows us to access their dynamics at strong coupling. On the other hand, they have been subject to recent scrutiny within the context of the Swampland Programme \cite{Vafa:2005ui,Brennan:2017rbf,Palti:2019pca,vanBeest:2021lhn,Grana:2021zvf}, where several proposals to describe their general properties have been made. Out of them, the most relevant one for the present work is the AdS Instability Conjecture \cite{Ooguri:2016pdq,Freivogel:2016qwc}, which states that all $\cN=0$ AdS$_d$ vacua are unstable, in which case their holographic description would not make much sense. In particular, in perturbatively stable vacua supported by $d$-form fluxes, the instability is expected to arise at the non-perturbative level, from one or several superextremal $(d-2)$-branes that nucleate and expand towards the AdS$_d$ boundary \cite{Maldacena:1998uz}. The existence of such branes is predicted by a sharpening of the Weak Gravity Conjecture (WGC), which states that the WGC inequality is only saturated in supersymmetric settings \cite{Ooguri:2016pdq}. 

All these statements are particularly meaningful in string constructions where the compactification scale is much smaller than the AdS length scale, as then the nucleation can be described by means of an EFT valid at intermediate scales. In this sense, the DGKT-CFI proposal \cite{DeWolfe:2005uu,Camara:2005dc}, in which a parametric separation of scales is achieved by moving in an infinite family of AdS$_4$ vacua, represents an interesting arena to test these ideas. A quite general construction realising this feature is based on massive type IIA string theory compactified on a Calabi--Yau orientifold geometry with O6-planes and D6-branes and threaded by background fluxes,\footnote{See \cite{Derendinger:2004jn,Villadoro:2005cu} for previous similar constructions in toroidal orbifold settings.} and it is typically referred to as DGKT-like vacua. While a holographic description of these vacua remains elusive and some of their features are quite counter-intuitive \cite{Lust:2019zwm}, the proposal has passed non-trivial tests at the gravity side, like the approximate 10d description provided in \cite{Junghans:2020acz,Marchesano:2020qvg}. 

A general classification of DGKT-like vacua can be done using a 4d EFT description, which includes an F-term potential generated by fluxes that fixes the Calabi--Yau moduli. Such an analysis was carried out in \cite{Marchesano:2019hfb}, where at least four branches of perturbatively stable vacua -- one supersymmetric and three non-supersymmetric -- were shown to exist. All of these branches contain an infinite number of vacua that is generated by a rescaling of internal fluxes, as in the supersymmetric case, and along which parametric scale separation is achieved. Remarkably, the mass spectrum found in \cite{Marchesano:2019hfb} for some of these branches has an amusing holographic interpretation \cite{Conlon:2021cjk,Apers:2022tfm}, while the remaining branches do not present this feature \cite{Quirant:2022fpn}. 

Given this setup, the purpose of this work is to gain further insight into the non-supersymmetric branches of DGKT-like vacua, and in particular \bk{on} their perturbative and non-perturbative stability, following up on previous work on this subject \cite{Junghans:2020acz,Marchesano:2021ycx,Casas:2022mnz}. As a first step, one would like to confirm the 4d result on perturbative stability, or in other words to verify that the F-term potential from where the moduli masses are derived is reliable. The effective F-term potential in massive type IIA orientifold compactifications used in \cite{Marchesano:2019hfb} is derived either by performing a direct Kaluza--Klein reduction over an Calabi--Yau geometry threaded by internal fluxes \cite{Grimm:2004ua,Grimm:2011dx,Kerstan:2011dy}, or through the formalism of 4d three-form potentials  \cite{Bielleman:2015ina,Carta:2016ynn,Farakos:2017jme,Herraez:2018vae,Bandos:2018gjp,Lanza:2019xxg}. If one obtains a 10d description for these vacua that displays scale separation and an approximate Calabi--Yau metric, then it means that the derivation of the potential is accurate up to the said degree of approximation. \bk{This was shown to be the case in \cite{Junghans:2020acz} via a general description of approximate solutions to 10d massive IIA equation that correspond to DGKT-like vacua.} The degree of accuracy is given by the 10d dilaton vev or equivalently by the inverse AdS$_4$ length in string units, which both become parametrically small as we advance in the infinite family of vacua. \bk{In the present work we confirm this picture by reproducing the four 4d branches of vacua mentioned above directly from a 10d perspective. The 10d background describing all these 4d vacua is provided at the same degree of explicitness as given for the supersymmetric branch in  \cite{Marchesano:2020qvg}, using a combination of the results in \cite{Junghans:2020acz} and  \cite{Marchesano:2020qvg}.}

We then turn to analyse the non-perturbative stability of these vacua, by considering the charge $Q$ and tension $T$ of their 4d membranes, along the lines of \cite{Aharony:2008wz,Narayan:2010em,Marchesano:2021ycx,Casas:2022mnz}. We focus in particular on $D(2p+2)$-branes wrapping $2p$-cycles of $X_6$ which are those that can nucleate in the context of the 4d EFT \cite{Lanza:2019xxg,Lanza:2020qmt}. According to the sharpened WGC at a generic vacuum one should find at least two membranes with $Q>T$. One made up of a D4-brane wrapping a two-cycle $\Sigma \subset X_6$ or a bound state containing it, and another one made up of a D8-brane wrapping $X_6$. Both objects we analysed in \cite{Marchesano:2021ycx} for \bk{one} branch of non-supersymmetric vacua, with special attention to the microscopic description of D8-branes as BIons. It was found that D4-branes satisfy $Q=T$ at the level of accuracy that we are working, while D8-branes satisfy $Q>T$ in simple configurations, due to a mixture of curvature corrections to their charge and tension and further corrections due to their BIonic nature. However, closer inspection showed that this last statement depends on the specific configuration of space-time filling D6-branes in a given vacuum, and that for some vacua the corrections to the D8-brane charge and tension tip the scales towards $Q<T$  \cite{Casas:2022mnz}. Therefore, it would seem that in such vacua not only the sharpened WGC fails to be true, but even the WGC for 4d membranes itself. 

As we will see, this apparent tension with the WGC is solved when one considers more exotic D-brane configurations. In particular, we look at those which are BPS  in supersymmetric DGKT vacua. Namely, we consider D8-branes wrapping $X_6$ and D6-branes wrapping a divisor ${\cal S} \subset X_6$, both threaded by non-diluted worldvolume fluxes in their internal dimensions (i.e., with worldvolume fluxes comparable to the K\"ahler two-form). One can check that at least one of these objects satisfies $Q\geq T$ in non-supersymmetric DGKT-like vacua. Since they couple to the same three-forms as D4-branes and D8-branes, they realise the WGC for 4d membranes. In fact, in most cases they correspond to superextremal 4d membranes, as predicted by the sharpened WGC. Only in one subclass of $\cN=0$ vacua all the relevant 4d membranes are extremal, namely in those $\cN=0$ vacua without space-time-filling D6-branes which, from the 4d viewpoint, are related to supersymmetric ones by an overall sign flip of the four-form flux. Quite amusingly, it is precisely such vacua which display integer conformal dimensions for their would-be holographic dual. Whether there is some meaning behind this coincidence or the marginality is an artefact of the accuracy of our description remains an open question for the future. 

The paper is organised as follows. In section \ref{s:branch} we review the main features of DGKT-like vacua and the four branches of solutions found  \cite{Marchesano:2019hfb}. In section \ref{s:10d} we discuss how to describe such 4d vacua from a 10d viewpoint, first using the smearing approximation and then with a more accurate 10d background with localised sources. In section \ref{s:membranes} we address the non-perturbative stability of these vacua by analysing the extremality of 4d membranes in the probe approximation. We leave our conclusions for section \ref{s:conclu} and several technicalities for the appendices. Appendix \ref{ap:10deom} analyses in detail the 10d equations of motion and Bianchi identities for all branches of vacua. Appendix \ref{ap:DBI} deduces the D-brane DBI expressions by means of which we compute the corresponding 4d membrane tension. 

\section{Branches of AdS$_4$ Calabi--Yau orientifold vacua}
\label{s:branch}

Our current understanding of  AdS$_4$ Calabi--Yau orientifold vacua is based on type IIA string theory compactified on a Calabi--Yau three-fold $X_6$. To this background we apply an orientifold quotient generated by $\Omega_p (-1)^{F_L}{\cal R}$,\footnote{Here $\Omega_p$ is the worldsheet parity reversal operator and ${F_L}$ is the space-time fermion number for the left-movers.} with ${\cal R}$ an anti-holomorphic involution of $X_6$ acting as ${\cal R} J_{\rm CY}=-J_{\rm CY}$, ${\cal R}\Omega_{\rm CY} = - \overline{\Omega}_{\rm CY}$ on its K\"ahler two-form and holomorphic three-form, respectively. The fixed locus $\Pi_{\rm O6}$ of ${\cal R}$ is made of  3-cycles of $X_6$, hosting O6-planes. The presence of O6-planes reduces the background supersymmetry to 4d $\CN=1$, and induces an RR tadpole that can be cancelled by a combination of D6-branes wrapping special Lagrangian three-cycles \cite{Blumenhagen:2005mu,Blumenhagen:2006ci,Marchesano:2007de,Ibanez:2012zz}, D8-branes wrapping coisotropic five-cycles \cite{Font:2006na}, and background fluxes including the Romans mass. If background fluxes are involved, one recovers a metric background of the form 
\begin{equation}\label{eq:warped-product}
	ds^2 = e^{2A}ds^2_{\mathrm{AdS}_4} + ds^2_{X_6}\, ,
\end{equation}
 with $A$ a function on $X_6$. This may either correspond to a 4d $\CN=1$ or $\CN=0$ vacuum. 

If O6-planes and background D-branes are treated as localised sources, the warping function $A$ is non-constant. Similarly, we have a 10d dilaton $e^\phi$ varying over $X_6$ with an average value $g_s$, and a metric on $X_6$ which is no longer Calabi--Yau, but should instead be a deformation to a $SU(3)\times SU(3)$ structure metric. This picture is based on the results of \cite{Junghans:2020acz,Marchesano:2020qvg}, which provided explicit approximate solutions for the 10d equations of motion and Bianchi identities of massive type IIA supergravity. Their key ingredient is an expansion of the said equations in a small parameter, which in the case at hand can be taken to be either $g_s$ or $|\hat{\mu}| = \ell_s/R$, the AdS$_4$ scale in 10d string frame and in string length $\ell_s  =  2\pi \sqrt{\a'}$ units \cite{Saracco:2012wc}. The zeroth order of the expansion treats $\delta$-function sources like O6-planes and D6-branes as if they were smeared over $X_6$, yielding a particularly simple solution with constant warping and dilaton, and a Calabi--Yau metric. The localised nature of these sources is already manifest in the first non-trivial correction to this background, which also displays the said deformation away from the Calabi--Yau metric.

The advantage of the smearing zeroth-order approximation is that it gives a direct connection with the 4d effective approach to describe these vacua. In the 4d picture one considers the set of moduli present in a large-volume Calabi--Yau compactification without fluxes, and a scalar potential generated by flux quanta that stabilises them at certain vevs. The 4d approach reveals an interesting vacua structure already in the case of toroidal orientifold compactifications \cite{Derendinger:2004jn,Villadoro:2005cu,DeWolfe:2005uu,Camara:2005dc}, and it can be generalised to arbitrary Calabi--Yau geometries thanks to the simple form of the scalar potential in the large-volume regime \cite{Herraez:2018vae,Escobar:2018tiu,Escobar:2018rna,Marchesano:2019hfb,Marchesano:2020uqz}. In the following we  review the results of \cite{Marchesano:2019hfb}, which obtained several branches of supersymmetric and non-supersymmetric vacua using the 4d approach on arbitrary Calabi--Yau orientifold geometries. 

Calabi--Yau orientifold vacua can be described by a set of relations between the Calabi--Yau metric forms $\Omega_{\rm CY}$ and $J_{\rm CY}$ and the  background fluxes. To describe the latter it is convenient to use the democratic formulation of type IIA supergravity \cite{Bergshoeff:2001pv}, in which all RR potentials are grouped in a polyform ${\bf C} = C_1 + C_3 + C_5 + C_7 + C_9$ and so are their gauge invariant field strengths
\be
{\bf G} \,=\, d_H{\bf C} + e^{B} \wedge {\bf \bar{G}} =  {\rm vol}_4 \wedge \tilde{G} + \hat{G} \, .
\label{bfG}
\ee
Here $H$ is the three-form NS flux, $d_H \equiv (d - H \wedge)$ is the $H$-twisted differential and ${\bf \bar{G}}$ a formal sum of closed $p$-forms on $X_6$. The second equality is specific to the metric background \eqref{eq:warped-product}, with ${\rm vol}_4$ the AdS$_4$ volume form, $\tilde{G}$ and $\hat{G}$ only have internal indices and satisfy the relation $\tilde{G} = - \lambda ( *_6 \hat{G})$, and where $\lambda$ is the operator that reverses the order of the indices of a $p$-form. 
The Bianchi identities for these field strengths read
\begin{equation}\label{IIABI}
\ell_s^{2} \,  d (e^{-B} \wedge {\bf G} ) = - \sum_\a \lambda \left[\delta (\Pi_\alpha)\right] \wedge e^{\frac{\ell_s^2}{2\pi} F_\alpha} \, ,  \qquad d H = 0 \, ,
\end{equation} 
where $\Pi_\alpha$ hosts a D-brane source with a quantised worldvolume flux $F_\alpha$, and $\delta(\Pi_\alpha)$ is the bump $\delta$-function form with support on $\Pi_\alpha$ and indices transverse to it, such that $\ell_s^{p-9} \d(\Pi_\a)$ lies in the Poincar\'e dual class to $[\Pi_\a]$. O6-planes contribute as D6-branes but with minus four times their charge and $F_\alpha \equiv 0$. In the absence of localised sources, each $p$-form within ${\bf \bar{G}}$ is quantised, so one can define the internal RR flux quanta in terms of the following integer numbers
\begin{equation}
m \, = \,  \ell_s G_0\, ,  \quad  m^a\, =\, \frac{1}{\ell_s^5} \int_{X_6} \bar{G}_2 \wedge \tilde \omega^a\, , \quad  e_a\, =\, - \frac{1}{\ell_s^5} \int_{X_6} \bar{G}_4 \wedge \omega_a \, , \quad e_0 \, =\, - \frac{1}{\ell_s^5} \int_{X_6} \bar{G}_6 \, ,
\label{RRfluxes}
\end{equation}
with $\omega_a$, $\tilde \omega^a$ integral Calabi--Yau-harmonic two- and four-forms such that $\ell_s^{-6} \int_{X_6} \omega_a \wedge \tilde{\omega}^b = \delta_a^b$, in terms of which we can expand the K\"ahler form as
\be
J_{\rm CY} = t^a \omega_a\, , \qquad - J_{\rm CY} \wedge J_{\rm CY} = {\cal K}_a \tilde{\omega}^a\, . 
\ee
Here ${\cal K}_a \equiv {\cal K}_{abc} t^bt^c$, with ${\cal K}_{abc} = - \ell_s^{-6} \int_{X_6} \omega_a \wedge \omega_b \wedge \omega_c$ the Calabi--Yau triple intersection numbers and $-\frac{1}{6} J_{\rm CY}^3 = - \frac{i}{8} \Omega_{\rm CY} \wedge \bar{\Omega}_{\rm CY}$ its volume form.

Even in the presence of localised sources, \eqref{RRfluxes} are taken as integer flux quanta that together with the H-flux quanta enter the F-term scalar potential. The latter has a series of extrema that have been classified in \cite{Marchesano:2019hfb}. In the following we consider four of the branches of vacua found therein, dubbed as class {\bf S1}. They consist of one infinite family of supersymmetric vacua and three non-supersymmetric ones. Given the 4d moduli stabilisation data, which in the conventions of this paper is reviewed in \cite[Appendix A]{Marchesano:2021ycx}, one obtains that the background fluxes describing such vacua must obey the following relations:
\be
[ H ]  = 6A G_0 g_s  [\re \Omega_{\rm CY} ] \, , \qquad \frac{1}{\ell_s^6} \int_{X_6} {G}_2 \wedge \tilde{\omega}^a =  BG_0 t^a\, ,  \qquad -\frac{1}{\ell_s^6} \int_{X_6} \hat{G}_4  \wedge \omega_a  =  CG_0 {\cal K}_a  \, , 
\label{intflux}
\ee
together with $\hat{G}_6  =  0$. Here $A, B, C \in \pr$ are constants that index the different branches, see table \ref{vacuresul} for their specific values. The stabilisation of Calabi--Yau moduli in terms of flux quanta follows from these relations and
\be
\hat{e}_a \equiv e_a - \oh \frac{\cK_{abc} m^bm^c}{m}  = \left( C - \oh B^2\right) m {\cal K}_a \, .
\label{hate}
\ee

An important feature of these vacua is that the quanta of $H$-flux and $G_0$ are constrained by the RR-flux Bianchi identities, that in the presence of O6-planes and D6-branes read
\be
dG_0 = 0\, , \qquad d G_2 = G_0 H - 4 \d_{\rm O6} +   N_\a \d_{\rm D6}^\a \, ,  \qquad d \hat{G}_4 = G_2 \wedge H\, , \qquad d\hat{G}_6 = 0\, ,
\label{BIG}
\ee
 we have defined $\d_{\rm D6/O6}\equiv \ell_s^{-2}  \d(\Pi_{\rm D6/O6})$ and $N_\a$ is the number D6-branes wrapping the three-cycle $\Pi^{\rm D6}_\a$. This in particular implies that
\be
{\rm P.D.} \left[4\Pi_{\rm O6}- N_\a \Pi_{\rm D6}^\a\right] = m [\ell_s^{-2} H]  \implies mh +N = 4 \, ,
\label{tadpole}
\ee
where to arrive to the last equation we have taken the simplifying choice P.D.$[\ell_s^{-2}H] = h [\Pi_{\rm O6}] = h [\Pi_{\rm D6}^\a]$, $\forall \a$. In all branches $A>0$, so it follows from \eqref{intflux} and that all sources are calibrated by $\im\, \Omega_{\rm CY}$ that $0 < mh \leq 4$. The remaining flux quanta $e_a, m^a$ are however unconstrained by RR tadpole conditions, and so one can choose them freely to fix ${\cal K}_a$ arbitrarily large. As one does, it is driven to a region of larger Calabi--Yau internal volume ${\cal V}_{\rm CY}  =  \frac{1}{6} {\cal K}_{abc}t^at^bt^c \equiv  \frac{1}{6} {\cal K}$, weaker 10d string coupling $g_s$ and smaller AdS$_4$ curvature. The latter is given by
\be
\mu = G_0 g_s \frac{2}{3}\sqrt{C^2 + \frac{1}{8}B^2}\, ,
\label{mu}
\ee
again measured in the 10d string frame. 

\begin{table}[H]
\begin{center}
\scalebox{1}{%
    \begin{tabular}{| c || c | c | c | c |c |c|}
    \hline
  Branch & $A$  & $B$  & $C$  & $\mu$  & SUSY & pert. stable \\
  \hline \hline
  \textbf{A1-S1}$+$  &$\frac{1}{15}$ &  $0$  & $\frac{3}{10}$  & $\frac{1}{5} G_0 g_s$ & Yes & Yes  \\ \hline
    \textbf{A1-S1}$-$  &$\frac{1}{15}$  & $0$  & $-\frac{3}{10}$  & $\frac{1}{5} G_0 g_s$ & No & Yes \\ 
\hline
     \textbf{A2-S1}$\pm$    & $\frac{1}{12}$  & $\pm\frac{1}{2}$  & $-\frac{1}{4}$  & $\frac{1}{\sqrt{24}} G_0 g_s$ 	& No & Yes \\ 
       \hline
    \end{tabular}}      
\end{center}
\caption{Different branches of {\bf S1} solutions found in  \cite{Marchesano:2019hfb}. \label{vacuresul}}
\end{table}

Table \ref{vacuresul} shows the four different branches of solutions found in  \cite{Marchesano:2019hfb} that correspond to the relations \eqref{intflux}, with the different values for  $A, B, C$. The branch \textbf{A1-S1}$+$ corresponds to the infinite family of supersymmetric solutions found in \cite{DeWolfe:2005uu}, while \textbf{A1-S1}$-$ represents non-supersymmetric vacua whose four-form flux harmonic piece has a sign flip compared to the supersymmetric case. Just like their supersymmetric cousins, these non-supersymmetric vacua have a simple, universal flux-induced mass spectrum absent of tachyons below the BF bound \cite{Marchesano:2019hfb}. Finally, the branches \textbf{A2-S1}$\pm$ correspond to non-supersymmetric vacua that have been less studied in the literature. While their mass spectrum is harder to analyse in general (see \cite{Quirant:2022fpn} for the case of toroidal geometries), one can show that the potential is positive semidefinite \cite{Marchesano:2019hfb}, and therefore they are perturbative stable as well.

Since they only differ by the value of the constants $A, B, C$, all these branches have the same parametric dependence on their Kaluza--Klein and AdS scales for larges values of $\hat{e}$. In particular they reproduce the scaling $m_{\rm KK} \sim \hat{e}^{1/2} \mu$ observed in  \cite{DeWolfe:2005uu} for the supersymmetric branch that leads to parametric scale separation. If this estimate of scales survives the 10d description of these vacua, it means that we can trust our 4d effective analysis, and in particular the perturbative stability obtained from it. In the next section we will address the 10d description of all these branches, extending the analysis of \cite{Marchesano:2020qvg,Marchesano:2021ycx}. We will see that from the smearing approximation one can rederive table \ref{vacuresul}, and then provide the first-order correction to this approximation, that describes localised sources. Since in principle this confirms the perturbative stability of such vacua, we turn to analyse their non-perturbative stability in section \ref{s:membranes}.


\section{10d uplift and localised sources}
\label{s:10d}

In this section we recover the 4d results reviewed above from a 10d viewpoint. As we will see, the four branches of {\bf S1} solutions can also be obtained by solving the equations of massive type IIA supergravity up to a certain order in a perturbative expansion, following \cite{Saracco:2012wc,Junghans:2020acz,Marchesano:2020qvg}. We first show that solving the equations at zeroth order, in which localised sources appear to be smeared, already reproduces the four different branches of table \ref{vacuresul}. We then proceed to show that the solution for each of these branches of vacua can be extended to the first order in the perturbative expansion, where space-time O6-planes and D6-branes are treated as localised sources  in the internal dimensions.

\subsection{Smearing approximation}

Let us first address the 10d equations of motion and Bianchi identities in the smearing approximation. Since we are not restricted to supersymmetric backgrounds, we will follow the general approach of \cite{Junghans:2020acz}. In such a formalism, after making a perturbative expansion of the 10d equations,  one obtains that the zeroth order 10d equations are described by a smearing approximation, which is defined by means of the following prescription:

\begin{itemize}

\item The metric on $X_6$ is taken to be Calabi--Yau, the warp factor dilaton to be constant, and the background fluxes to have a harmonic $p$-form profile in this metric. This implies that the flux Ansatz \eqref{intflux} is approximated by the following, more specific flux background
\be
H   = 6A G_0 g_s  \re \Omega_{\rm CY}  \, , \qquad  {G}_2 =  BG_0 J_{\rm CY}\, ,  \qquad  \hat{G}_4   =  CG_0  J_{\rm CY} \wedge  J_{\rm CY}  \, , \qquad\hat{G}_6  =  0\, . 
\label{intfluxsm}
\ee

\item The three-form bump $\delta$-functions that appear in the Bianchi identities \eqref{BIG} are replaced by harmonic representatives in the same homology class. Taking in addition the simplifying choice of eq.\eqref{tadpole} one obtains that the only non-trivial Bianchi identity at this level reads
\be
 d G_2 = G_0 H - mh \delta_{\rm O6}^{\rm h} = 0\, ,
 \label{BIG2sm}
\ee
where $ \delta_{\rm O6}^{\rm h}$ is the harmonic piece of the three-form bump $\delta$-function $\delta_{\rm O6}$. 

\item The $\delta$-like sources $\delta_\a^{(3)}$ that appear in the dilaton and Einstein equations are replaced by constant terms describing their zero mode in a Fourier expansion. Assuming that all three-cycles wrapped O6-plane and D6-brane are calibrated by $\Im \Om$, as we will do in the following, one can relate these localised sources with the three-form bump functions as 
\be
\delta^{(3)}_{\a} \equiv *_{6} (\im \Omega \wedge \delta(\Pi_{\a})) \simeq *_{\rm CY} (\im \Omega_{\rm CY} \wedge \delta(\Pi_{\a})) \to \frac{{\cal V}_{\Pi_\a}}{{\cal V}_{\rm CY}}\, ,
\label{deltasm}
\ee
where in the second step we have taken the Calabi--Yau metric approximation and in the third one we have replaced the $\delta$-function by its zero mode. Here ${\cal V}_{\Pi_\a}$ is the volume of the three-cycle $\Pi_\a$ measured in string units. 

\end{itemize}

Applying these prescriptions to the 10d massive type IIA supergravity equations, one obtains a set of constraints on the parameters $A, B, C \in \pr$. In particular, from the Bianchi identity \eqref{BIG2sm} one obtains
\be
\frac{mh}{\ell_s^{2}}\frac{{\cal V}_{\Pi_{\rm O6}}}{{\cal V}_{\rm CY}} = 24AG_0^2 g_s \, ,
\label{BIG2c}
\ee
where we have also made use of \eqref{deltasm}. Additionally, by plugging \eqref{intfluxsm} into the equations of motion for the background fluxes one obtains that the only non-trivial equation is 
\be
\label{fluxeomsm}
{G}_2 \wedge  *_{\rm CY} \hat{G}_4  + {G}_0 *_{\rm CY} {G}_2  = - G_0^2\,  J_{\rm CY} \wedge  J_{\rm CY}\, B\left(2C+ \oh\right)\, ,
\ee
see Appendix \ref{ap:10deom} for details. Solving this equation already constrains the parameters of our Ansatz, to either satisfy $B=0$ or $C=-1/4$. Notice that, in the language of \cite{Marchesano:2019hfb}, these two choices precisely correspond to the branches {\bf A1} and {\bf A2}, respectively.

Finally, one must apply the above prescription to the 10d Einstein and dilaton equations of motion. One obtains the following relations
\bes
\label{10dEinsteinsm}
\begin{align}
\label{10ddilsm}
\mu^2 & = \frac{G_0^2g_s^2}{72} \left( 144A^2 +3B^2 +36C^2-1\right)\, , \\
\label{10dE1sm}
\frac{mh}{\ell_s^{2}}\frac{{\cal V}_{\Pi_{\rm O6}}}{{\cal V}_{\rm CY}} & = \frac{G_0^2g_s}{3} \left( 576A^2 +3B^2 +36C^2-1\right)\, , \\
\label{10dE2sm}
\frac{mh}{\ell_s^{2}}\frac{{\cal V}_{\Pi_{\rm O6}}}{{\cal V}_{\rm CY}} & = \frac{G_0^2g_s}{6} \left( 1584A^2 +3B^2 +84C^2-5\right)\, . 
\end{align}
\ees
From \eqref{10dE1sm} and  \eqref{10dE2sm} one finds 
\be
144A^2-1 = B^2 -4C^2\, ,
\ee
which plugged into \eqref{10ddilsm} reproduces \eqref{mu}. Additionally, using \eqref{BIG2c} and \eqref{10dEinsteinsm} one obtains that
\be
72A = 3+7B^2+20C^2\, .
\ee
These last two equations and \eqref{fluxeomsm} completely determine the allowed values for the parameters of our flux Ansatz. For the branch {\bf A1} one recovers $A =1/15$ and $C=\pm3/10$, while for the branch {\bf A2} one finds $A=1/12$ and $B =\pm 1/2$, precisely reproducing the content of table \ref{vacuresul}.

\subsection{First-order corrections and localisation}

Let us now proceed beyond the smearing approximation and solve the 10d equations at the next order in the $g_s$ expansion. For this we follow the same strategy as in \cite{Marchesano:2021ycx}, and combine the results of \cite{Junghans:2020acz} and \cite{Marchesano:2020qvg}. More precisely, we consider the same metric and dilaton background obtained in \cite{Marchesano:2020qvg} for the supersymmetric case, and then we apply the approach in \cite{Junghans:2020acz} to obtain the flux background that solves the 10d equations at the same order of approximation. 

In the first-order solution found in \cite{Marchesano:2020qvg}, the background corresponding to \eqref{eq:warped-product} is described by a $SU(3)\times SU(3)$ structure metric on $X_6$ and a varying dilaton and warp factor of the form
\begin{subequations}	
	\label{solutionsu3}
\begin{align}
J & = J_{\rm CY} + \cO(g_s^2) \, , \qquad   \qquad  \Omega  = \Omega_{\rm CY} + g_s k +  \cO(g_s^2)\, , \\
e^{-A}  & = 1 + g_s \varphi + \cO(g_s^2) \, , \qquad e^{\phi}   = g_s \left(1 - 3  g_s \varphi\right) + \cO(g_s^3)\, ,
\end{align}
\end{subequations}   
where $k$ is a (2,1) primitive current  and $\varphi$ a real function that satisfies $\int_{X_6} \varphi = 0$. These two quantities are obtained by solving the Bianchi identity \eqref{BIG} for $G_2$ at the given order of approximation in the $g_s$ expansion. Expressing the internal two-form flux as
\be
G_2 = G_2^{\rm h} + d^\dag_{\rm CY} K + \cO(g_s)\, ,
\ee
where $G_2^{\rm h}$ is given by the smeared profile in \eqref{intfluxsm}, and $K$ is three-form current satisfying
\begin{equation}
    \Delta_{\rm CY} K = G_0H +  \delta_{\rm O6+D6}    = 6A G_0^2 g_s  \re \Omega_{\rm CY} -  mh \delta_{\rm O6}  +  \cO(g_s^2)\, ,
    \label{eq: K equation}
\end{equation}
where we have defined $\Delta_{\rm CY} = d^\dag_{\rm CY} d + d d^\dag_{\rm CY}$ and used \eqref{tadpole} and \eqref{intfluxsm}. The harmonic piece of the RHS of this equation vanishes due to \eqref{tadpole}, or equivalently due to \eqref{BIG2sm}. Hence there is always a solution for $K$, which at this order of approximation is of the form
\be
 K = \varphi \re \Omega_{\rm CY}  + \re k \, ,
\label{formK}
\ee
with $\varphi$ satisfying a Laplace equation with $\delta$-sources on top of the O6-planes and D6-branes, see \cite{Casas:2022mnz} for a detailed discussion and several explicit examples. 

Given the above metric, dilaton and two-form flux background one may look for the profiles of the remaining internal fluxes such that {\it i)} they reduce to the smeared values \eqref{intfluxsm} at the lowest order in the $g_s$ expansion and {\it ii)} they solve the 10d equations of massive type IIA supergravity at the next order in the same expansion. This exercise is carried out in Appendix \ref{ap:10deom}, with the following result
\begin{subequations}
	\label{solutionflux}
\begin{align}
 H & =   6A G_0 g_s \left(\re \Omega_{\rm CY} + R g_s K \right) -\frac{S}{2}   d\re \left(\bar{v} \cdot \Omega_{\rm CY} \right) + \cO(g_s^{3}) \label{H3sol} \, , \\
 \label{G2sol}
 G_2 & =    BG_0 J_{\rm CY} - J_{\rm CY} \cdot d(4 \varphi \im \Omega_{\rm CY} - \star_{\rm CY} K) + \cO(g_s) \, , \\
G_4 & =   G_0 J_{\rm CY} \wedge J_{\rm CY} \left(C  - 12A g_s \varphi \right)+  S J_{\rm CY} \wedge g_s^{-1} d \im v + \cO(g_s^2) \, , \\
G_6 & = 0\, ,
\end{align}
\end{subequations}   
where $R, S \in \pr$ and $v$ is a (1,0)-form determined by
\be \label{ansatzVF}
v  = g_s \p_{\rm CY} f_\star + \cO(g_s^3)\, , \qquad \text{with} \qquad \Delta_{\rm CY} f_\star  = - g_s 8 G_0 \varphi \, .
\ee 
This background has the same form as in the supersymmetric case, and only differs \bk{on} the values that the constants $A,B,C,R,S$ take, which are different for each branch. The value of the new constants $R$ and $S$ are in fact determined by those that already appear in the smearing approximation, as follows
\be
6AR = 12A + 2C -1\, , \qquad S = 6A + 2C\, ,
\ee
yielding the content of table \ref{vacuresulns}.
\begin{table}[H]
\begin{center}
\scalebox{1}{%
    \begin{tabular}{| c || 	c | c | c | c |c |c|c|c|}
    \hline
  Branch & $A$  & $B$  & $C$  & $R$ & $S$ & $\mu$  & SUSY & pert. stable \\
  \hline \hline
  \textbf{A1-S1}$+$  &$\frac{1}{15}$ &  $0$  & $\frac{3}{10}$  & $1$ & $1$ & $\frac{1}{5} G_0 g_s$ & Yes & Yes  \\ \hline
    \textbf{A1-S1}$-$  &$\frac{1}{15}$  & $0$  & $-\frac{3}{10}$  & $-2$ & $-\frac{1}{5}$ & $\frac{1}{5} G_0 g_s$ & No & Yes \\ 
\hline
     \textbf{A2-S1}$\pm$   & $\frac{1}{12}$  & $\pm\frac{1}{2}$  & $-\frac{1}{4}$ &$-1$ & $0$ & $\frac{1}{\sqrt{24}} G_0 g_s$ 	& No & Yes \\ 
       \hline
    \end{tabular}}      
\end{center}
\caption{Different branches of {\bf S1} solutions found in  \cite{Marchesano:2019hfb}, beyond the smearing approximation. \label{vacuresulns}}
\end{table}

These results suggest that we can have a 10d description for each of the vacua in table \ref{vacuresul} in which the internal geometry is well approximated by the Calabi--Yau metric, which is more and more accurate for larger values of $\hat{e}$, hence smaller values of $g_s$. In such a regime, our 4d estimate for the Kaluza--Klein scale is accurate, and below it we can trust our 4d effective potential, including the values for the flux-induced moduli masses. As a result, our 10d backgrounds should be free of perturbative instabilities, including those which belong to non-supersymmetric branches. It however remains to analyse their non-perturbative decay channels, and in particular those mediated by nucleating 4d membranes, which we now turn to discuss.


\section{4d membranes and non-perturbative instabilities}
\label{s:membranes}

To detect non-perturbative instabilities of the vacuum triggered by membrane nucleation one may follow \cite{Aharony:2008wz,Narayan:2010em} and consider probe 4d membranes that extend along a hyperplane $z = z_0$ within the Poincar\'e patch of AdS$_4$
\be
ds^2_4 =e^{\frac{2z}{R}} (-dt^2 + d\vec{x}^2) + dz^2\, ,
\label{PPatch}
\ee
where $R= |\mu|^{-1}$ is the AdS length scale, $\vec{x} = (x^1, x^2)$, and all coordinates range over $\pr$. A membrane with non-trivial tension $T$ will naturally be dragged towards $z \to -\infty$, except if it couples as $-\int C_3$ to a background four-form flux with vev $Q$
\be
\langle F_4 \rangle = -\frac{3Q}{R} {\rm vol}_4
\qquad \Longrightarrow \qquad \langle C_3 \rangle = Q\, e^{\frac{3z}{R}} dt \wedge dx^1 \wedge dx^2 \, .
\label{3form}
\ee
We can interpret $Q$ as the  membrane charge with respect to a normalised three-form potential. 
Whenever $Q=T$  the energy dependence on $z_0$  due to the membrane tension cancels out with the potential energy $-\int \langle C_3 \rangle$ due to its charge. Moving the membrane along the coordinate $z$ is then a flat direction, as expected for BPS membranes. In fact, as argued in \cite{Koerber:2007jb}, membranes of this sort with $Q=T$ and near the AdS$_4$ boundary $z_0 \to \infty$ capture the BPS bound of spherical membranes in global coordinates at asymptotically large radius. This is particularly relevant for the stability of the vacuum, since it is precisely the domain walls that correspond to spherical membranes near the AdS boundary that determine whether the non-perturbative decay of one vacuum to another with lower energy is favourable or not. In this sense, one may interpret a membrane with $Q=T$ as mediating a marginal decay as it happens between supersymmetric vacua, while one with $Q>T$ is likely to signal a non-perturbative instability of the vacuum.\footnote{This correspondence typically assumes a thin domain wall, which is not always a good approximation.} 

Interestingly, the Weak Gravity Conjecture applied to 4d membranes implies that at each vacuum there must be one membrane with $Q \geq T$, for each independent membrane charge. Moreover, the refinement made in \cite{Ooguri:2016pdq} proposes that this inequality is only saturated in supersymmetric vacua. In non-supersymmetric vacua there should be a membrane with $Q>T$ for each independent membrane charge, therefore signalling an instability. In this section we consider these proposals in the context of the AdS$_4$ orientifold vacua of section \ref{s:branch}, following the same strategy as in \cite{Marchesano:2021ycx}, namely computing $Q$ and $T$ via dimensional reduction of D-brane actions. As we will see, the key observation to satisfy the WGC for 4d membranes is to consider branes whose internal dimensions are threaded by non-diluted worldvolume fluxes. 

\subsection{4d membrane charges and their Weak Gravity Conjecture}

In order to check the WGC and its refinement for the DGKT-like vacua of section \ref{s:branch}, let us start by reviewing and extending the results of \cite{Marchesano:2021ycx}, which addressed this question for {\bf A1-S1} vacua. First of all, one should make precise the WGC statement, in the sense that one should describe the set of independent membrane charges in these vacua. Naively, one would associate the set of membrane charges with the lattice of fluxes, as described by the $H$-flux and RR flux quanta \eqref{RRfluxes}. 
However, some of the points in this lattice do not correspond to independent flux quanta, as they are related to each other by large gauge transformations involving periodic shifts of the axions \`a la axion monodromy, see e.g. \cite{Berasaluce-Gonzalez:2012awn,Marchesano:2014mla,Herraez:2018vae}. After such identifications one is left with a set of membranes with torsional charges, that are related to discrete three-form gauge symmetries  \cite{Berasaluce-Gonzalez:2012awn,Buratti:2020kda}. It is not clear if the WGC should apply to such torsional membrane charges, but in the following we will not consider them. Instead, we will focus on those 4d fluxes that do not couple to any axion. In general, one can describe their quanta by using the set of flux invariants defined in \cite{Marchesano:2020uqz}. In DGKT-like vacua, such invariants reduce to the $H$-flux quanta, the Romans' parameter $m$ and the flux combinations $\hat{e}_a$. Hence, in a given vacuum one should look for membranes that, as one crosses them towards $z\to -\infty$, they make one of these flux quanta jump and take us to a vacuum with lower energy, or equivalently with larger AdS$_4$ scale $|\mu|$. In practice this amounts to jumps that decrease $|\hat{e}_a|$ and/or increase $|m|$ or $|h|$. Notice that the last two are constrained by the tadpole condition \eqref{tadpole}, and so in some cases it is not possible to increase their value. In those cases only the membranes that change $|\hat{e}_a|$ should be considered. 

As in \cite{Marchesano:2021ycx}, we only consider those 4d membranes that arise from wrapping D$(2p+2)$-branes on $2p$-cycles of $X_6$. Such membranes couple to the dynamical fluxes of the 4d theory \cite{Lanza:2019xxg,Lanza:2020qmt}, which include the flux quanta $m$ and $\hat{e}_a$ and exclude the $H$-flux quanta. The charge of each of these membranes can be obtained by dimensionally reducing their Chern-Simons action, which couples to the appropriate component of the flux polyform $\tilde{G}$ defined in \eqref{bfG}. Similarly to \cite{Marchesano:2021ycx}, one finds that in the smearing approximation the different charges read
\be
Q_{\rm D2} = 0 \, , \qquad Q_{\rm D4} =  \frac{ C }{D} \eta q_{\rm D4} T_{\rm D4}  \, , \qquad Q_{\rm D6} =  \frac{B}{2D} \eta q_{\rm D6}   T_{\rm D6} \, , \qquad Q_{\rm D8} =  -\frac{\eta q_{\rm D8}}{2D} T_{\rm D8}\, ,
\label{QDGKT}
\ee
where we have assumed that a D4-branes wraps a holomorphic curve $\Sigma$, a D6-brane a divisor ${\cal S}$, and D8-branes the whole of $X_6$, so their tension in 4d Planck units is given by 
\be
T_{\rm D2} = 1 \, , \qquad T_{\rm D4} = e^{K/2}{\cal V}_\Sigma \, , \qquad T_{\rm D6} =  e^{K/2}{\cal V}_{\cal S} \, , \qquad T_{\rm D8} =  e^{K/2}{\cal V}_{\rm CY}\, ,
\label{TDGKT}
\ee
with ${\cal V}$  the volume of each cycle in string units. Their orientation of the cycle, or equivalently if we consider a D-brane or an anti-D-brane, is encoded in $q_{{\rm D}(2p+2)} =\pm1$. Finally we have defined
\be
\eta = {\rm sign}\, m\, , \qquad D= \sqrt{C^2 + \frac{1}{8}B^2}\, .
\ee
It is easy to see that these results reproduce those in sections 3 and 4 of \cite{Marchesano:2021ycx}. In there the branches {\bf A1-S1} were considered, for which $B=0$ and so $Q_{\rm D4} = \eta \eta_C q_{\rm D4}  T_{\rm D4}$, with $\eta_C = {\rm sign}\, C$. One just needs to choose $ q_{\rm D4}$ such that $\eta \eta_C q_{\rm D4} = 1$ for the extremal condition $Q=T$ to be met. As expected,  this choice corresponds to the D4-branes that decrease the value of $|\hat{e}_a|$ \cite{Marchesano:2021ycx}. In particular, the case $m >0$ selects D4-branes for supersymmetric {\bf A1-S1}$+$ vacua and anti-D4-branes for the non-supersymmetric branch {\bf A1-S1}$-$. The opposite choice leads to $Q=-T$. 

As also pointed out in \cite{Marchesano:2021ycx}, the energetics of D8-branes is more involved that for the rest, because they have an excess of space-time-filling D6-branes ending on them and stretching along $z \in [z_0, \infty)$ for $\eta  q_{\rm D8} = 1$, and along  $z \in (-\infty, z_0]$ for $\eta  q_{\rm D8} = -1$. Their presence contributes to the forces acting on the D8-brane transverse position, so that it can be encoded in an effective D8-brane charge. Generalising the computations in  \cite{Marchesano:2021ycx} one finds that
\be
Q_{\rm D8}^{\rm eff} =  \frac{24A-1}{2D} \eta q_{\rm D8} T_{\rm D8}\, ,
\label{QD8eff}
\ee
which for {\bf A1-S1} reduces to $Q_{\rm D8}^{\rm eff} =  \eta q_{\rm D8} T_{\rm D8}$. Therefore, by taking $q_{\rm D8} = \eta$, which corresponds to a flux jump that increases $|m|$, one finds again a marginal membrane jump. 

To sum up, for {\bf A1-S1} vacua one finds that 4d membranes made up from both D4-branes and D8-branes satisfy $Q=T$, at least when computing these quantities in the smearing approximation. This is expected for {\bf A1-S1}$+$ vacua, which are supersymmetric, but would contradict the refinement of the WGC for the non-supersymmetric {\bf A1-S1}$-$ vacua. In order to check such a refinement one should then consider corrections to the 4d membrane charge and tension. Just like for the 10d background, such corrections can be expanded in increasing powers of $g_s$. For the case of D4-branes, one may look at corrections to $Q$ and $T$ that come from considering the more precise metric and flux backgrounds \eqref{solutionsu3} and \eqref{solutionflux}. It turns out that such corrections vanish for both classes of {\bf A1-S1} vacua, and so D4-branes wrapping (anti-)holomorphic two-cycles yield extremal 4d membranes also for {\bf A1-S1}$-$ vacua, at least at this level of the approximation. 

The story for D8-branes wrapping $X_6$ is slightly more involved \cite{Marchesano:2021ycx}. First, their DBI and CS actions are subject to curvature corrections encoded in the second Chern class of $X_6$, such that they can be understood as a bound state of a D8-brane and {\em minus} a D4-brane wrapping the Poincar\'e dual of $c_2(X_6)/24$. The term minus refers to the fact that these curvature corrections induce negative D4-brane and tension. This does not affect the relation $Q=T$ in supersymmetric vacua, but it changes it towards $Q>T$ for non-supersymmetric  {\bf A1-S1}$-$ vacua, due to the sign flip for the internal four-form flux $\hat{G}_4$. This provides a mechanism analogous to the one pointed out in \cite{Maldacena:1998uz}, where a D5-branes wraps the $K3$ in AdS$_3 \times S^3 \times K3$, and which drags the resulting membrane towards the AdS boundary. 

However, such curvature corrections appear at the same order in $g_s$ as the first corrections to the smearing approximation, and so both effects should be considered simultaneously. For D8-branes, corrections due to source localisation appear in two different ways. On the one hand, due to considering their DBI+CS action in the more precise background \eqref{solutionsu3} and \eqref{solutionflux}. On the other hand, by realising that the space-time-filling D6-branes ending on them are also localised sources for their worldvolume flux $\cF = B + \frac{\ell_s^2}{2\pi} F$. This second effect results in a BIon profile along the D8-brane transverse direction $z$, that encodes the energy of the D8/D6-brane system. Taking all the localisation effects into account one obtains a correction to the quantity $Q-T$ in  {\bf A1-S1}$-$ vacua of the form $2\Delta_{\rm D8}^{\rm Bion} \equiv - e^{K/2} \frac{1}{\ell_s^6} \int_{\rm X_6}  J_{\rm CY} \wedge \cF^2_{\rm BIon}$, where $\cF_{\rm BIon}$ is the piece of D8-brane worldvolume flux sourced by the  the D6-branes ending on it  and the $H$-flux \cite{Marchesano:2021ycx}. This quantity was computed in \cite{Casas:2022mnz} for several toroidal orbifold geometries, where it was compared to the D8-brane curvature corrections. It was found that $\Delta_{\rm D8}^{\rm Bion}$ can have both signs depending on the relative positions of the space-time-filling D6-branes in such vacua. In particular, it was found that in some instances adding both sets of corrections tips the scale towards $Q<T$, in apparent tension with the (unrefined) WGC for 4d membranes. 

Despite these negative results, in the following we will argue that the WGC for membranes is satisfied in the DGKT-like vacua of section \ref{s:branch}. To do so, we will consider more exotic bound states of D$(2p+2)$-branes, and in particular D8 and D6-branes with non-diluted worldvolume fluxes threading their internal dimensions. The corresponding 4d membranes will not only provide new decay channels for {\bf A1-S1}$-$ vacua, but also for the non-supersymmetric branches {\bf A2-S1}. Indeed, notice that for the latter $D = \sqrt{3/32}$, and so the ratio $Q/T$ for a 4d membrane obtained from wrapping a plain D$(2p+2)$-brane is given by an irrational number smaller than one. Again, it is via considering exotic bound states that one can achieve 4d membranes with $Q>T$. 

\subsection{Exotic bound states of membranes}

Let us consider new D-brane bound states that are candidates to yield 4d membranes with $Q>T$ in non-supersymmetric vacua. The main strategy will be to identify those bound states that yield $Q=T$ in the supersymmetric case, and analyse similar objects in the non-supersymmetric branches. As we will see, the mismatch between $Q$ and $T$ arises at level of the smearing approximation, so we may phrase most of our discussion in terms of the approximate Calabi--Yau geometry. As advanced, the bound states of interests correspond to D8 and D6-branes with non-diluted worldvolume fluxes in the internal dimensions. More precisely, in the smeared approximation they can be described by the following conditions
 \bes
\label{exoticBPS}
\begin{align}
\label{exoticD8}
\text{D8-brane on $X_6$:} & \qquad \cF \wedge \cF = 3 J_{\rm CY} \wedge J_{\rm CY}\, ,\\
\label{exoticD6}
\text{$k$ D6-branes on ${\cal S}$:} & \qquad \cF \wedge \cF = J_{\rm CY} \wedge J_{\rm CY}|_{\cal S}\, .
\end{align}
\ees
Here  $\cF = B + \frac{\ell_s^2}{2\pi} F$ is the worldvolume flux\footnote{Recall that in the smearing approximation $\cF_{\rm Bion}=0$, so also for D8-branes $\cF$ is a closed two-form.} threading the internal dimensions of the D$(2p+2)$-brane, which is the whole $X_6$ in the case of D8-branes and a divisor ${\cal S}$ in the case of D6-branes. 

We dub these objects exotic bound states because, in the large volume regime, they carry a  large lower-dimensional D-brane charge, induced by a large flux $\cF$ \cite{Douglas:1995bn}.\footnote{One should not confuse the two notions of charge present in our discussion. D-brane charges refer to the couplings of D$(2p+2)$-branes to the RR $(2p+1)$-form potentials in 10d supergravity, in the absence of background fluxes. The charges in \eqref{QDGKT} correspond instead to the 4d membrane charges \eqref{3form} obtained via dimensional reduction of a Chern-Simons action in a particular 10d flux background that corresponds to a vacuum.} This makes them exotic from the  model building viewpoint, as parametrically large D-brane charges can be in conflict with RR tadpole conditions. 
 In the case at hand, the large D-brane charges carried by these bound states translate into 4d membranes that induce large shifts for the flux quanta $m^a$, $e_a$ and $e_0$, which are not constrained by tadpole conditions. Therefore,  one must consider them as part of the spectrum of 4d membranes, and as such they may mediate decays in non-supersymmetric vacua. We will now analyse their properties in the different DGKT-like branches of section \ref{s:branch}.

\subsubsection*{Supersymmetric vacua}

Let us discuss the properties of the D-branes \eqref{exoticBPS} in supersymmetric vacua. In fact, it proves useful to first consider the case of type IIA Calabi--Yau orientifold compactifications to Minkowski, in the absence of background fluxes. In this context, \eqref{exoticBPS} are particular solutions to the MMMS equations \cite{Marino:1999af} and as such the corresponding D-branes are BPS objects. One can also detect the BPSness of such D-branes by 
analysing their DBI action, see Appendix \ref{ap:DBI}. After imposing \eqref{exoticBPS} the DBI action linearises and its integrand reads\footnote{Curvature corrections will modify this expression as well as the BPS conditions \eqref{exoticBPS}, shifting $\cF \wedge \cF$ by   $c_2(X_6)/24$ for D8-branes and by $c_2({\cal S})/24$ for D6-branes. Because this effect is subleading in the large volume regime, and is comparable to corrections to the smearing approximation, it will be neglected in the following. }
\be
d {\rm DBI} = e^{i \theta} g_s^{-1} \left.e^{-(\cF + iJ_{\rm CY})} \right|_{2p}  
\label{calibration}
\ee
where $p=3$ for D8-branes and $p=2$ for D6-branes. We say that both objects are calibrated by $e^{-(\cF + iJ_{\rm CY})}$, with $e^{-i\theta}$ their calibration phase. In compactifications to Minkowski 4d membranes are BPS for any calibration phase, but only two membranes with the same calibration phase are mutually BPS. This is a relevant statement because, as we will show below, in $\CN=1$ AdS$_4$ type IIA vacua all 4d membranes that are BPS have the same calibration phase. We have already run into some BPS 4d membranes in supersymmetric DGKT-like vacua, like a D4-brane on a holomorphic curve $\Sigma$, which corresponds to $\theta = \pi/2$. Other D-branes with the same phase are
\bes
\label{iBPS}
\begin{align}
\label{iD6}
\text{(anti-)D6-brane on ${\cal S}$ with} & \quad \cF^2 = J_{\rm CY}^2|_{\cal S}\, , \\
\label{iD8}
\text{D8-brane on $X_6$ with} & \quad  \cF^2 \wedge J_{\rm CY} = c J^3_{\rm CY} \, , \ c \leq 0 \quad {\rm and}  \quad 3\cF \wedge J_{\rm CY}^2 = \cF^3 \, , \\
\label{iaD8}
 \text{anti-D8-brane on $X_6$ with} & \quad \cF^2 = 3 J_{\rm CY}^2\, .
\end{align}
\ees
We have encountered instances of \eqref{iD8} in our previous discussion, like the case $c=0$ which corresponds to a D8-brane with $\cF = 0$. Other cases in which $0 > c \sim \cO({\cal V}_{\rm CY}^{-2/3})$ represent D8-branes with a worldvolume flux that is approximately primitive, and corresponds to a solution to the $\alpha'$-corrected Donalson-Uhlenbeck-Yau equations \cite{Douglas:2001hw}. Such objects can be seen as bound states of D8-branes and a few D6, D4 and D2-branes, and were also considered as 4d membranes in \cite{Marchesano:2021ycx}. However, they are not particularly interesting from the viewpoint of the WGC for 4d membranes in non-supersymmetric vacua. On the one hand they carry positive D4-brane charge, and in {\bf A1-S1}$-$ vacua this contributes towards $Q<T$. So in order to look for membranes with $Q \geq T$ it is better to set $c = 0$. On the other hand, they are quite unnatural in {\bf A2-S1} vacua, because the non-diluted B-field sets $c \sim \cO(1)$. In any event, we see that our reasoning selects two new candidates for 4d membranes satisfying the WGC, which are quite similar to \eqref{exoticBPS}.

Let us now show that all these objects fulfil the extremal condition $Q=T$ in supersymmetric AdS$_4$ vacua. For this, we consider the 10d type IIA  supersymmetry conditions \cite[eq.(2.13)]{Marchesano:2020qvg} in the smearing approximation
\bes
\label{susy}
\begin{align}
\label{susyg}
d_H \im \Omega_{\rm CY} & -  g_s *_6 \left(G_0 - G_2 + \hat{G}_4 - \hat{G}_6 \right) + 3 \mu \im \left( e^{-iJ_{\rm CY}}\right) = 0\, , \\
d_H e^{-iJ_{\rm CY}} & + 2\mu \re \Omega_{\rm CY} = 0\, .
\label{susydw}
\end{align}
\ees
We may pull-back \eqref{susyg} on a $2p$-cycle of $X_6$ wrapped by a D$(2p+2)$-brane. Then, by multiplying the result by $e^{-\cF}$ and using \eqref{bfG} one obtains
\be
\frac{1}{3\mu} \left(g_s^{-1} d_H\Omega_{\rm CY} - e^{-\cF} \wedge \tilde{G}\right)_{2p} = - g_s^{-1}  \im \left( e^{-\cF-iJ_{\rm CY}}\right)_{2p}  = \sin \theta\, d {\rm DBI}\, ,
\label{10dQT}
\ee
where in the last equality we have used \eqref{calibration}. By switching off the worldvolume flux $\cF$, one can see that both sides of this equation are related to the 4d membrane charge and tension that appear in \eqref{QDGKT}. By introducing $\cF$ one generalises this notion for bound states that arise from such worldvolume fluxes. Indeed, upon integration of the rhs of \eqref{10dQT} one recovers $\sin \theta$ times the 4d membrane tension. Similarly, the lhs of \eqref{10dQT} encodes the 4d membrane effective charge. Upon integration on an internal $2p$-cycle it gives $\eta Q$, where $\eta = {\rm sign}\, m$ and
\be
Q = \frac{\eta e^{K/2}}{\ell_s^{2p}}  \int_{2p}  e^{-\cF} \wedge {\bf Q} \, ,  \quad \text{with} \quad  {\bf Q} = \sum_{p}  \frac{q_p}{p!}J_{\rm CY}^p\, ,
\label{Qform}
\ee
and the coefficients $q_p$ correspond to charge-to-tension ratios $Q_{{\rm D}(2p+2)}/T_{{\rm D}(2p+2)}$. Namely,
\be
q_0 = 0\, , \qquad q_1 = \frac{C}{D}\, , \qquad q_2 = -\frac{B}{2D}\, ,  \qquad q_3 = -\frac{24A-1}{2D}\, . 
\ee
Note that $Q$ reproduces \eqref{QDGKT} and \eqref{QD8eff} for 4d membranes with $\cF=0$ in all branches of DGKT-like vacua, and it extends the definition of charge to their bound states. In the supersymmetric branch we have that ${\bf Q} =  \im e^{iJ_{\rm CY}}$, and \eqref{10dQT} translates into 
\be
Q = \eta \sin \theta\, T\, .
\ee
This illustrates our claim that, in supersymmetric AdS$_4$ vacua, all 4d membranes with $Q=T$ have the same calibration phase. In the case at hand they have $\theta = \pi/2$ for $m>0$, like D4-branes wrapping holomorphic curves $\Sigma$ and the D-branes in \eqref{iBPS}. For $m<0$ they must instead have $\theta=-\pi/2$, like anti-D4-branes on $\Sigma$ and the anti-D-branes version of \eqref{iBPS}. It is easy to convince oneself that our reasoning is more general that the specifics of DGKT-like vacua, and it ultimately boils down to the interpretation of the 10d supersymmetry equations as the existence of calibrations for D-branes wrapping internal cycles of a compact manifold \cite{Martucci:2005ht,Koerber:2007jb}.

It may seem surprising that anti-D8-branes with worldvolume fluxes in supersymmetric vacua with $m>0$ can be BPS and that their transverse position in the AdS$_4$ coordinate $z$ is a flat direction. In this case, the D8-brane tension and (effective) charge add up to drag them away from the AdS boundary. However, the worldvolume flux condition \eqref{iaD8} implies that they form a bound state with a very large number $N \sim 9T_{\rm D8}/T_{\rm D4}$ of D4-branes. Hence, even if D4-branes have smaller tension, their large number makes them weight nine times more than a D8-brane. The bound state is calibrated by $\frac{4}{3} J_{\rm CY}^3$, which given the opposite orientation compared to $X_6$ results in a 4d membrane tension  $T= 8 T_{\rm D8}$, while $Q= -T_{\rm D8} + NT_{\rm D4} =  8 T_{\rm D8}$. It thus happens that the tension gained by the bound state as compared to its constituents precisely cancels the factor of $2T_{\rm D8}$ that would drag away from the AdS boundary an anti-D8-brane with $\cF=0$. 

Before turning to non-supersymmetric vacua, let us comment on the actual existence of the D-branes \eqref{iD6} and \eqref{iaD8} in supersymmetric DGKT vacua. The question is non-trivial, because in such vacua the K\"ahler moduli and the B-field axions take discrete values as a function of the background flux quanta. So everything is fixed in these BPS equations except the piece of worldvolume flux given by $F$, which is also quantised. Hence, for arbitrary values of the complexified K\"ahler moduli one may not be able to find examples of such D-branes, which also illustrates the somewhat exotic nature of these objects.

Let us first consider the anti-D8-branes. Since they wrap the whole of $X_6$, the equation in \eqref{iaD8} is directly related to the stabilisation of Calabi--Yau moduli. In particular, \eqref{hate} implies
\be
3 J_{\rm CY}^2 = -10 \left(\frac{e_a}{m} - \oh \frac{\CK_{abc}m^bm^c}{m^2}\right) \tilde{\omega}^a \, ,
\ee
while the stabilisation of B-field axions implies that
\be
\cF =  \left(n^a -\frac{m^a}{m} \right) \omega_a\, ,
\ee
where $n^a \in \pz$. Putting both conditions together one finds that \eqref{iaD8} amounts to
\be
\CK_{abc} \left(m^2n^bn^c - 2n^bm^c\right)  = 10 m e_a - 6 \CK_{abc}m^bm^c\, ,  \quad \forall a\, .
\label{dioD8BPS}
\ee
Given some choice of flux quanta $m, m^a, e_a$, one should find appropriate values of $n^a$ solving these equations. While both sides of \eqref{dioD8BPS}  are integer, it is not always true that a solution to such quadratic Diophantine equations exist. In the particularly simple case where $m^a = 0$ they reduce to $m\CK_{abc} n^bn^c = 10 e_a$, which do not have a solution unless $10e_a$ is a multiple of $m$, $\forall a$. 

Similar equations can be derived for the case of D6-branes. Assuming $k$ D6-branes wrapped on a Nef divisor ${\cal S}_a$ dual to $\omega_a$, and a quantised worldvolume flux of the form $\frac{\ell_s^2}{2\pi} F = \frac{n^b}{k} \mathbbm{1}_{k} \omega_b|_{{\cal S}_a}$, the BPS condition \eqref{iD6} amounts to 
\be
\CK_{abc} \left(\frac{mn^bn^c}{k} - 2 n^bm^c \right)  = \frac{10k}{3} e_a - \frac{8k}{3m} \CK_{abc}m^bm^c \, ,
\label{dioD6BPS}
\ee
where $\CK_{abc}n^bn^c/k \in \IZ$ must be satisfied \cite{Rabadan:2001mt}. In this case we have a single Diophantine equation to solve, and we have more freedom, in the sense that given $m, m^a, e_a$ we may adjust the values of both $k$ and $n^a$ to find solutions. In particular, it seems that one must take $k$ proportional to $3m$ in order to find solutions for generic values of the flux quanta. In the particular case where $m^a = 0$, the equation reduces to $\CK_{abc} \frac{n^bn^c}{k} = \frac{10k}{3m}e_a$, which should have solution whenever $[e_a\tilde{\omega}^a]$ is dual to the intersection of two divisors.

\subsubsection*{A1-S1$-$ vacua}

Let us now turn to non-supersymmetric {\bf A1-S1}$-$ vacua. Recall that these vacua are defined by a sign flip of the internal four-form $\hat{G}_4$ with respect to the supersymmetric ones. In other words, $Q_{\rm D4}$ flips sign and (for $m>0$) D4-branes wrapping holomorphic curves satisfy $Q=-T$ from the 4d viewpoint, while anti-D4-branes satisfy $Q=T$. For this reason, 4d membranes corresponding to \eqref{iaD8} cannot satisfy $Q>T$ in {\bf A1-S1}$-$ vacua with $m >0$, since both of their constituents (anti-D8-brane and D4-branes) contribute with a negative charge. One may instead consider their anti-object, which is nothing but \eqref{exoticD8}. As we will see, the corresponding 4d membrane satisfies $Q>T$ and provides a decay channel for this class of vacua. 

Indeed, \eqref{exoticD8} can be roughly seen as a bound state of a D8-brane and $N \sim 9T_{\rm D8}/T_{\rm D4}$ anti-D4-branes. The charges of both constituents are positive whenever $m>0$, and add up to $Q = 10 T_{\rm D8}$. Indeed, for D8-branes in {\bf A1-S1}$-$ vacua we have that \eqref{Qform} reads
\be
Q =  \frac{\eta e^{K/2}}{\ell_s^{6}}  \int_{X_6}  e^{-\cF} \wedge \left( -  J_{\rm CY} - \frac{1}{6} J_{\rm CY}^3\right) = 10 \eta T_{\rm D8} \, ,
\label{QA1S1-}
\ee
where in the second equality we have applied \eqref{exoticD8}. The 4d membrane tension is equal to that of its anti-object \eqref{iaD8}, namely $T = 8 T_{\rm D8}$, as can also be checked by using the results of appendix \ref{ap:DBI}. Therefore we obtain that $Q-T = 2T_{\rm D8}$, and the membrane is superextremal. 

It remains to see whether these objects actually exist for a given vacuum. As in their supersymmetric case, their BPS condition translates into a quadratic Diophantine equation:
\be
\CK_{abc} \left(mn^b -m^b\right)\left(mn^c -m^c \right) = -10 m\hat{e}_a\, ,
\ee
or equivalently
\be
\CK_{abc} \left(m^2n^bn^c - 2n^bm^c\right)  = - 10 m e_a + 4 \CK_{abc}m^bm^c\, ,  \quad \forall a\, .
\label{dioD8A1-}
\ee
These equations look a bit different from the supersymmetric case, but note from \eqref{hate} that  ${\rm sign}\, (m\hat{e}_a) = \pm$ for {\bf A1-S1}$\pm$ vacua, so we are essentially solving the same equations. As before, we do not expect that for arbitrary choices of $m, m^a, e_a$ one can find $n^a \in \pz$ such that all these equations are satisfied. In that case, D8-branes with $\cF^2 = 3 J_{\rm CY}^2$ do not exists. However, one can still argue that D8-branes with worldvolume fluxes such that $Q>T$ do still exist. 

Indeed, let us consider that the quantised piece of the worldvolume flux is of the form
\be
\frac{\ell_s^2}{2\pi} F = n^a \omega_a \, , \qquad \text{with} \qquad n^a = \pm\sqrt{3} t^a + \frac{m^a}{m} + \eps^a\, . 
\label{quantansatz}
\ee
Here $\eps^a \in \pr$ are chosen to be the smallest possible numbers such that $n^a \in \pz$ and $\eps^a \hat{e}_a = 0$. Generically, this second condition sets $\eps^a$ to be of the order of the largest quotient between two $\hat{e}_a$'s, which we denote by $M$. It also implies that we can write the worldvolume flux as
\be
\cF = \pm \sqrt{3} J_{\rm CY} + \cF_{\rm p}\, ,
\ee
where $\cF_{\rm p} = \eps^a \omega_a$ is a primitive (1,1)-form, that is $\cF_{\rm p} \wedge J_{\rm CY}^2 =0$. Plugging this expression for the worldvolume flux into eq.\eqref{ap:DBID8}, one obtains that the D8-brane DBI density reads
\be
d{\rm DBI}_{\rm D8} = g_s^{-1} \sqrt{\left(8 -  ||\eps||^2 \right)^2 + \left(\sqrt{3} ||\eps||^2 + \cO(||\eps||^3) \right)^2} d{\rm vol}_{X_6}\, ,
\label{DBID8eps}
\ee
where we have defined
\be
||\eps|| = \frac{1}{2} \sqrt{\cF_{{\rm p}, ab} \cF_{\rm p}^{ab} }   \sim \cO\left(\frac{M}{{\cal V}_{\rm CY}^{1/3}}\right)\, .
\label{epsdef}
\ee
In the following we will assume that $||\eps|| \ll 1$, because it is not clear that otherwise we have scale separation, or even that the K\"ahler moduli are stabilised in the supergravity regime. Under this assumption one can expand \eqref{DBID8eps} and obtains that the tension reads
\be
T = \left(8 -  ||\eps||^2_0 + 2  ||\eps||^4_0 + \dots\right) T_{\rm D8}\, ,
\ee
where we have defined $||\eps||^n_0 \equiv \int_{X_6} ||\eps||^n/ {\cal V}_{\rm CY}$, and the dots represent higher order terms in $||\eps||$. Similarly, one may compute the membrane tension from \eqref{QA1S1-}, obtaining
\be
Q = \left(10 -  ||\eps||^2_0 \right) \eta T_{\rm D8}\, .
\ee
Therefore we find that for $m > 0$
\be
Q - T =  2\left(1 -  ||\eps||^4_0 + \dots\right) T_{\rm D8}\, ,
\ee
and the membrane is superextremal. For $m<0$, one instead needs to consider the anti-D8-brane \eqref{iaD8} to find the same result. 

Even if these 4d membranes may not be in the thin-wall approximation, one may apply the reasoning of \cite[section 5]{Marchesano:2021ycx} to argue that they represent a non-perturbative instability towards a vacuum with larger $|m|$ and smaller $|\hat{e}|$. Still they should not be considered in vacua where $|m|$ cannot be made larger due to the tadpole constraints, as for instance in models without space-time-filling D6-branes like in \cite{DeWolfe:2005uu}. In those cases, only membranes that vary $m^a$, $e_a$ and $e_0$ should be considered, like D4-branes and the D6-branes in \eqref{exoticD6}. For the latter, and under the same assumptions as in \eqref{dioD6BPS} their existence translates in the following  Diophantine equation
\be
\CK_{abc} \left(\frac{mn^bn^c}{k} - 2 n^bm^c \right)  = - \frac{10k}{3} e_a + \frac{2k}{3m} \CK_{abc}m^bm^c \, ,
\label{dioD6A1-}
\ee
which is again quite similar to that of the supersymmetric branch. As in there, we expect that one can choose appropriate values of $k$ and $n^a$ to find a solution. In the smearing approximation, we have that the tension of the corresponding 4d membrane is similar to its supersymmetric counterpart. Using eq.\eqref{ap:DBID6} and applying \eqref{exoticD6} one finds
\be
T = e^{K/2} \frac{1}{\ell_s^4} \left| \int_{\cal S} \cF \wedge J_{\rm CY}  \right| \, 
\ee
while its charge can be computed via \eqref{Qform} 
\be
Q =  e^{K/2} \frac{\eta }{\ell_s^{4}}  \int_{\cal S}  \cF \wedge  J_{\rm CY}\, .
\ee
The sign of the integral will depend of the sign of the projection of $\cF$ into $J_{\rm CY}|_{\cal S}$. For either sign and for each value of $\eta$ one can satisfy the extremal condition $Q=T$ by either considering a D6-branes or an anti-D6-brane satisfying \eqref{exoticD6}. 

We therefore only find superextremal 4d membranes when they arise from D8-branes. D4-branes and the D6-branes \eqref{exoticD6} are at best marginal. As this would contradict the WGC refinement proposed in \cite{Ooguri:2016pdq}, one may wonder if the equality $Q=T$ is an artefact of the smearing approximation. Following the same computations as in \cite[section 6]{Marchesano:2021ycx}, one can convince oneself that the D6-brane charge and tension do not vary when we consider them in the more precise background \eqref{solutionsu3} and \eqref{solutionflux}. Finally, one may add curvature corrections to the D6-brane action, which will modify its tension. However, the same corrections will also modify the worldvolume flux condition \eqref{exoticBPS}, in such a way that both effects cancel out. Therefore, at the level of approximation that we are working, we find that DGKT-like vacua in the {\bf A1-S1}$-$ branch without space-time-filling D6-branes are marginally stable. Whether further corrections tip the scale  towards $Q>T$ or not remains an open problem.

\subsubsection*{A2-S1 vacua}

Let us consider the last two branches of vacua, namely {\bf A2-S1}$\pm$, which can be discussed simultaneously. In this case, one can also show that D-brane bound states \eqref{exoticBPS} lead to 4d membranes with $Q>T$ whenever they exist. Discussing their existence is however more involved than in the \bk{\bf A1-S1}$\pm$ branches. Indeed, in the present vacua the worldvolume flux of, say, a D8-brane is of the form 
\be
\cF =  \left(n^a + Bt^a - \frac{m^a}{m} \right) \omega_a\, ,
\ee
and so when plugged into \eqref{exoticD8} there will be an explicit dependence on the K\"ahler moduli. As such, it is difficult to determine whether such an equation has a solution, unless the vevs of the K\"ahler moduli are known explicitly as a function of the background fluxes. 

\begin{figure}[H]
    	\centering
		\includegraphics[width=0.65\linewidth]{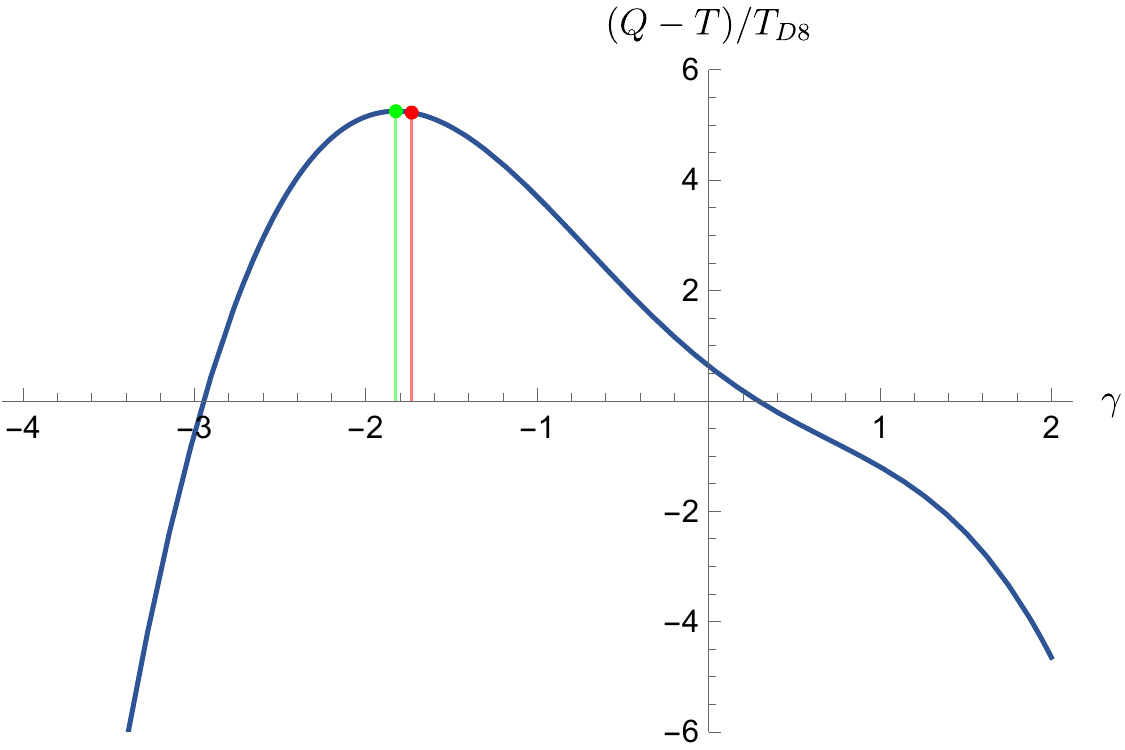}
		\caption{\label{plotD8} $Q-T$ for D8-branes in units of $T_{\rm D8}$ over $\gamma$ (blue) with $\eta = \eta_B = 1$ and $\epsilon = 0$. The dots correspond to the maximum of the curve $\gamma \simeq -1.82$ (green), and to $\gamma = - \sqrt{3}$ (red). $Q>T$ for the range $-2.95 \lesssim \gamma \lesssim 0.29$.}
\end{figure}

Nevertheless, one may still implement the approach previously used for D8-branes, and consider that the quantised piece of the worldvolume flux is of the form 
\be
\frac{\ell_s^2}{2\pi} F = n^a \omega_a \, , \qquad \text{with} \qquad n^a = \left(\gamma - B\right) t^a + \frac{m^a}{m} + \eps^a\, ,
\label{quantansatzA2}
\ee
with $\gamma \in \pr$ and $\eps^a$ satisfying the same constraints as in \eqref{quantansatz}. The corresponding worldvolume flux reads
\be
\cF = \gamma J_{\rm CY} + \cF_{\rm p}\, ,
\label{cfg}
\ee
and one may compute the 4d membrane tension and charge in terms of its parameters. As before, by plugging \eqref{cfg} into eq.\eqref{ap:DBID8} one obtains the following DBI density for D8-branes
\be
d{\rm DBI}_{\rm D8} = g_s^{-1}\sqrt{\left(3\g^2 - 1 -  ||\eps||^2 \right)^2 + \left(\g \left( \g^2 -3  - ||\eps||^2\right) + \cO(||\eps||^3) \right)^2} d{\rm vol}_{X_6}\, ,
\label{DBID8A2}
\ee
where $||\eps||$ is defined as in \eqref{epsdef}. From here one deduces that the corresponding 4d membrane tension reads
\be
T = \left[\left(1+\gamma^2\right)^{3/2} -  \left(\g^4-1\right) ||\eps||^2_0  + \dots\right] T_{\rm D8}\, .
\ee
The 4d membrane charge can be computed by plugging \eqref{cfg} into \eqref{Qform}:
\be
Q =  \frac{\eta e^{K/2}}{\ell_s^{6}}   \sqrt{\frac{2}{3}} \int_{X_6}  e^{-\cF} \wedge \left( - J_{\rm CY} - \eta_B \oh J_{\rm CY}^2 - \frac{1}{3} J_{\rm CY}^3\right) = \sqrt{\frac{2}{3}} \left( 3\g^2 - 3 \g \eta_B  +2 -  ||\eps||^2_0\right)  \eta T_{\rm D8} \, ,
\label{QA2S1}
\ee
where we have defined $\eta_B \equiv {\rm sign}\, B$.  From these expressions it is easy to see that $Q>T$ for $\gamma=-\eta_B \sqrt{3}$ and $||\eps||\ll1$, as claimed above. However, this value of $\gamma$ does not give the maximum possible value of $Q-T$. The actual value of the maximum and the range for which $Q-T$ is positive can be evaluated numerically (see figure \ref{plotD8}).

\begin{figure}[H]
   	\centering
		\includegraphics[width=0.65\linewidth]{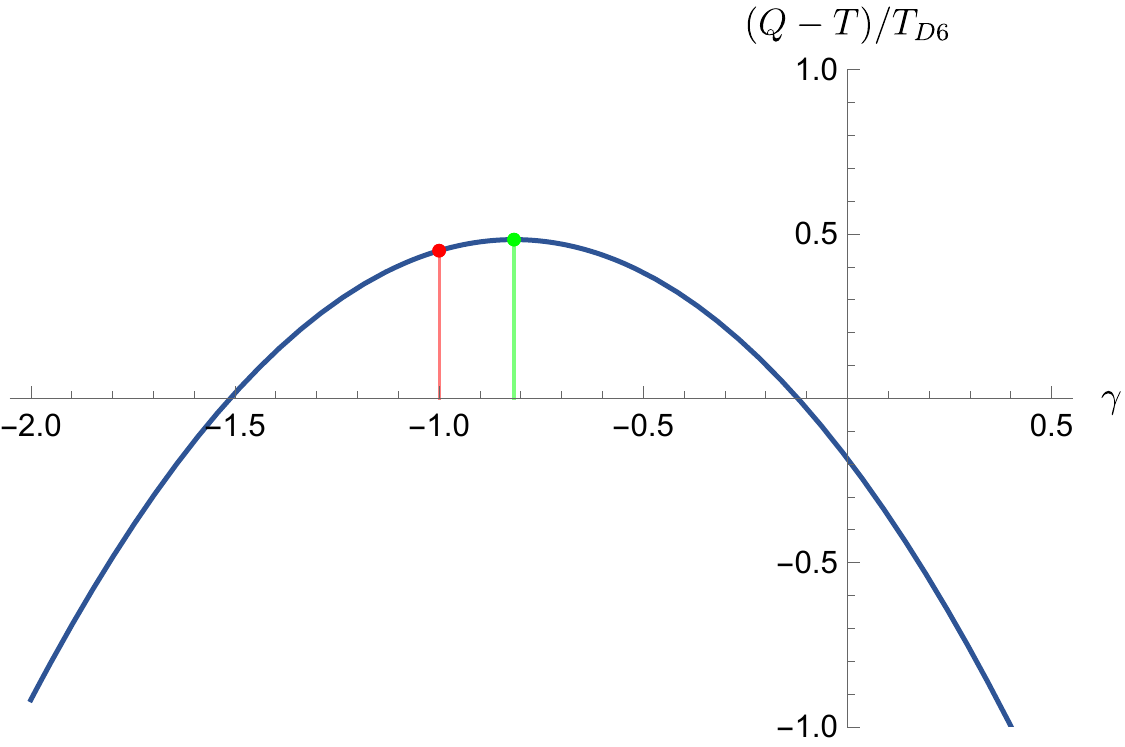}
			\caption{ \label{plotD6} $Q-T$ for D6-branes in units of $T_{D6}$ over $\gamma$ (blue) with $\eta = \eta_B = 1$ and $\epsilon = 0$ . The dots correspond to the maximum of the curve $\gamma = -\sqrt{2/3}$ (green) and to $\gamma = - 1$ (red). $Q>T$ for the range $-1.51 \lesssim \gamma \lesssim -0.12$.}
\end{figure}

One can implement the same strategy to analyse D6-branes with non-diluted worldvolume fluxes. This time, we take an Ansatz of the form \eqref{cfg} with $J_{\rm CY}$ representing the K\"ahler form pulled-back on the divisor ${\cal S}$, and $\cF_{\rm p}$ being a primitive (1,1)-form on the divisor, so that $\cF_{\rm p} \wedge J_{\rm CY} = 0$. We then encounter the following DBI density
\be
d{\rm DBI}_{\rm D6} = g_s^{-1} \sqrt{\left(\g^2 - 1 -  \varepsilon \right)^2 + 4\g^2}\, d{\rm vol}_{\cal S}\, ,
\label{DBID6A2}
\ee
where $\oh \cF_{\rm p} \wedge \cF_{\rm p} = \varepsilon\, d{\rm vol}_{\cal S}$. This leads to 
\be
T = \left(  1 + \g^2  -  \left(\g^2-1\right) \varepsilon_0  + \dots\right) T_{\rm D6}\, ,
\ee
with $\varepsilon_0 =  \int_{\cal S} \oh \cF_{\rm p} \wedge \cF_{\rm p}/ {\cal V}_{\cal S}$. The 4d membrane charge is again computed from \eqref{Qform}
\be
Q = \sqrt{\frac{2}{3}} \left(\eta_B-2\g \right) \eta T_{\rm D6}\, .
\ee
By choosing $\g = - \eta_B = - \eta$ one obtains that $Q>T$. Again, this is not the value of $\gamma$ that maximises $Q-T$. The actual value of the maximum and the range for which $Q-T$ is positive can be evaluated numerically (see figure \ref{plotD6}).


\section{Conclusions}
\label{s:conclu}

In this work we have analysed the perturbative and non-perturbative stability of DGKT-like vacua, following up on previous similar work \cite{Marchesano:2021ycx,Casas:2022mnz}. The vacua that can be built from a given Calabi--Yau manifold organise themselves on different branches, one of which is supersymmetric and the rest is non-supersymmetric. Out of the latter, three share some key properties with the supersymmetric branch, like an infinite set of vacua indexed by internal fluxes, parametric scale separation as we move along this set, and perturbative stability for all of them. While there are  obvious differences between each of these branches, we have managed to give a unified treatment for all of them in terms of their stability. The final result is summarised in table \ref{t:final}.
\begin{table}[H]
\begin{center}
\scalebox{1}{%
    \begin{tabular}{| c || c | c | c | c |c |c|}
    \hline
  Branch & SUSY & pert. stable & sWGC D4 & sWGC D8 & non-pert. stable  \\
  \hline \hline
  \textbf{A1-S1}$+$   & Yes & Yes  & Yes  & Yes & Yes \\ \hline
    \textbf{A1-S1}$-$  & No & Yes & Marginal  & Yes & unclear if $N_{\rm D6} =0$ \\ 
\hline
     \textbf{A2-S1}$\pm$   & No & Yes & Yes & Yes & No \\ 
       \hline
    \end{tabular}}      
\end{center}
\caption{Different branches of vacua, in terms of the sharpened WGC for membranes and their stability. \label{t:final}}
\end{table}

Perturbative stability can be addressed by providing a solution to the 10d equations of motion and Bianchi identities of massive type IIA that correspond to each of 4d vacua. The result is in section \ref{s:10d} and it follows the same approach of \cite{Junghans:2020acz,Marchesano:2020qvg}, by which one expands the 10d equations as a perturbative series in a small parameter (in our case $g_s$ or $\hat{\mu}$) and solves them up to next-to-leading order. The leading order corresponds to the so-called smearing approximation, which is the one used to derive the effective  potential used in \cite{Marchesano:2019hfb} to obtain a perturbatively stable vacuum for all the branches above. One can easily check that $\cO(g_s)$ corrections to the spectra derived in \cite{Marchesano:2019hfb} will not generate perturbative instabilities. 

The analysis of non-perturbative instabilities is more easily phrased in terms of the Weak Gravity Conjecture for 4d membranes, and more precisely via  the sharpening  proposed in \cite{Ooguri:2016pdq}. According to this more recent proposal,  there should be some extremal 4d membranes in  supersymmetric vacua, while non-supersymmetric vacua should contain superextremal membranes. There should be one of such objects per each independent membrane charge in our 4d EFT, which for the vacua of table \ref{t:final} translates into 4d membranes obtained from D4-branes and from D8-branes, or equivalently bound states that involve them. In terms of the definition of membrane charge $Q$ and tension $T$ given in the main text, in the supersymmetric branch \textbf{A1-S1}$+$ one should find membranes with $Q=T$, which is trivially satisfied by all D-branes that are (mutually) BPS in such a background. As for the non-supersymmetric branches, there should be at least one 4d membrane satisfying $Q>T$, separately for D4-branes and D8-branes. While this strict inequality is not aways realised by the most obvious choice of D8-branes \cite{Casas:2022mnz}, we have shown that by considering D8-branes threaded by non-diluted worldvolume fluxes one can construct 4d membranes that satisfy $Q>T$ in all $\cN=0$ branches, and therefore indicate an instability. Similarly, D6-branes with large internal worldvolume fluxes are bound states that involve D4-branes and which, in \textbf{A2-S1}$\pm$ vacua, correspond to membranes with $Q>T$, in line with the proposal in \cite{Ooguri:2016pdq}. The only case that escapes that proposal in the context of our analysis are D4-branes in \textbf{A1-S1}$-$ vacua, or any bound state without D8-brane charge. As already pointed out in \cite{Marchesano:2021ycx}, these objects are extremal even when considering first-order corrections to the smearing approximation. It could be that further corrections implement the inequality $Q>T$, but at the level of accuracy that we are working one should take some of the vacua in the \textbf{A1-S1}$-$ branch as marginally stable. In particular those vacua where the quantum of Romans mass $|m|$ cannot increase its value due to the tadpole constraint \eqref{tadpole}, like for instance when there are no space-time-filling D6-branes. 

In view of these results, it seems that a better understanding of DGKT-like vacua and their non-perturbative stability, as well as their connection with several Swampland criteria, seems like an interesting challenge for the future. We find particularly amusing that those non-supersymmetric vacua whose stability is still unclear at the current level of accuracy are those whose would-be holographic duals display integer conformal dimensions. Even if this coincidence does not seem to occur for 3d analogues \cite{Apers:2022zjx}, there could still something to be learnt if the same pattern is reproduced in further instances of 4d vacua. We hope that a more exhaustive analysis of the Landscape of AdS  vacua will shed some light into all these questions. 

\bigskip

\bigskip

\centerline{\bf  Acknowledgements}

\vspace*{.5cm}

We would like to thank David Prieto for useful discussions.  This  work is supported by the Spanish Research Agency (Agencia Estatal de Investigaci\'on) through the grants CEX2020-001007-S and PGC2018-095976-B-C21, funded by MCIN/AEI/10.13039/501100011033 and by ERDF A way of making Europe. J.Q. is supported through the FPU grant No. FPU17/04293 funded by MCIN/AEI/10.13039/501100011033 and by ESF Investing in your future. M.Z. is supported by the fellowship LCF/BQ/DI20/11780035 from ``la Caixa''
Foundation (ID 100010434).


\appendix


\section{10d equations of motion}
\label{ap:10deom}

In this appendix we analyse and solve the 10d equations of motion (EOMs) and Bianchi identities of massive type IIA supergravity, with a compactification Ansatz of the form \eqref{eq:warped-product}, using the conventions of \cite{Tomasiello:2022dwe}. 
 The Bianchi identities \ref{IIABI} for the polyform $\mathbf{G}$ encode both the EOMs and the Bianchi identities of the RR internal fluxes $\hat{G}$. Exploiting (\ref{bfG}) we obtain the Bianchi identities
\begin{subequations}\label{App10dEOMbianchi}
	\begin{align}
		& d{G}_0 = 0\,, \label{BiG0}\\
		& d {G}_2 = {G}_0 \wedge H - 4 \d_{\rm O6} +   N_\a \d_{\rm D6}^\a \, , \label{BiG2}\\
		& d \hat{G}_4 = {G}_2 \wedge H\, ,\label{BiG4} \\
		& d\hat{G}_6 = 0\, , \label{BiG6}
	\end{align}
\end{subequations}
and dualising the relations obtained for the external fluxes trough $\tilde{G} = -\lambda(\star_6 \hat{G})$ and the Ansatz \ref{eq:warped-product} we obtain the EOMs
\begin{subequations}
	\begin{align}
		& d(e^{4A}\star_{6}{G}_2) + e^{4A} H\wedge \star_{6}\hat{G}_4 = 0  \,, \label{eomG2} \\
		& 	d(e^{4A}\star_{6}\hat{G}_4) + e^{4A} H\wedge \star_{6}\hat{G}_6 = 0 \,, \label{eomG4}\\
		& 	d(e^{4A}\star_{6}\hat{G}_6) = 0 \label{eomG6} \,.
	\end{align}
\end{subequations}
The Bianchi identity and the EOM for the NSNS flux are instead 
\begin{subequations}
	\begin{align}
		& dH = 0 \,, \label{BiH} \\
		& d(e^{-2\phi+4A}\star_{6}H) + e^{4A}\star_{6} \hat{G}_6 \wedge \hat{G}_4 + e^{4A}\star_{6}\hat{G}_4 \wedge {G}_2 + e^{4A}\star_{6} {G}_2 \wedge {G}_0 = 0 \,.  \label{eomH} 
	\end{align}
\end{subequations}
The dilaton and the Einstein EOMs in our conventions are finally (see \cite{Junghans:2020acz})
\allowdisplaybreaks
\begin{subequations}
	\begin{align}
		\begin{split}
			0 = & \; 12 \frac{\tau^2}{\omega^2} + 12 \frac{\tau^2}{\omega^2}\left(\partial \omega \right)^2 + 4 \frac{\tau^2}{\omega} \nabla^2 \omega + 12 \frac{\tau}{\omega}\left(\partial \omega \right)\left(\partial \tau \right) + \tau \nabla^2 \tau 
			+ \left(\partial \tau \right)^2  - \frac{1}{2}\tau^2 \left|H\right|^2 \\ & \; - \sum_{q=0}^6 \frac{q-1}{4}|\hat{G}_q|^2  + \frac{\tau}{4}\sum \delta_i^{(3)} \,,
		\end{split} \\
		\begin{split}
			0 = & \; -\tau^2 R_{mn} + 4 \frac{\tau^2}{\omega} \nabla_m \partial_n \omega + \frac{\tau}{\omega}g_{mn}\left(\partial \omega\right)\left(\partial \tau \right) + \frac{1}{4}g_{mn}\tau \nabla^2 \tau + \frac{1}{4}g_{mn}\left(\partial \tau \right)^2 \\
			& \; + 2 \tau \nabla_m \partial_n \tau - 2\left(\partial_m \tau \right)\left(\partial_n \tau \right) 
			+ \frac{1}{2}\tau^2 \left(|H|_{mn}^2-\frac{1}{4}g_{mn}|H|^2\right) \\ & \; + \frac{1}{2}\sum_{q=0}^6\left(|\hat{G}_q|^2_{mn}- \frac{q-1}{8}g_{mn}|\hat{G}_q|^2\right) + \frac{\tau}{2}\sum_i \left(\Pi_{i,mn}-\frac{7}{8}g_{mn}\right)\delta_i^{(3)} \,,
		\end{split} \\
		\begin{split}
			0 = & \; -8 \nabla^2 \,\tau - 24 \frac{\tau}{\omega^2}-\frac{32}{\omega}\left(\partial \omega\right)\left(\partial \tau\right) - 24\frac{\tau}{\omega^2}\left(\partial \omega\right)^2 - 16 \frac{\tau}{\omega}\nabla^2 \omega + 2 \tau R_{mn}g^{mn} \\
			& \;  - \tau|H|^2 + \sum_i \delta^{(3)}_i\,,
		\end{split}
	\end{align}
\end{subequations}
with
\begin{equation}
	\Pi_{i,mn} = -\frac{2}{\sqrt{g_{\pi_i}}}\frac{\delta \sqrt{g_{\pi_i}}}{\delta g^{mn}} \,, \qquad |F_p|^2_{mn} = \frac{\delta |F_p|^2}{\delta g^{mn}}\,, \qquad \tau = e^{-\phi} \,, \qquad \omega = R e^A\,,
\end{equation}
with $R$ the AdS$_4$ radius, which implies that $\langle e^A \rangle = 1$. We finally have the delta-function sources
\begin{equation}
	\sum_i  \delta^{(3)}_i = \star_{\rm CY} \bigg[ \text{Im}_{\rm CY}\wedge \left(4 \d_{\rm O6} -   N_\a \d_{\rm D6}^\a\right) \bigg] \,.
\end{equation}

\subsection{Smearing approximation}

The smearing approximation assumes the dilaton and the warp factor to be constant, and the internal metric to be Calabi--Yau. With the Ansatz (\ref{intfluxsm}) and the tadpole cancellation condition (\ref{tadpole}) all fluxes are harmonic and the only non-trivial Bianchi identity is that of ${G}_2$ 
\begin{equation}
	d{G}_2 =  6 A  g_s {G}_0^2 \, \text{Re}(\Omega_{\rm CY}) - \frac{mh}{\ell_s^2} \delta_{O6} = 0  \,,
\end{equation}
which provides the following constraint
\begin{equation}\label{const1}
	24 A g_s {G}_0^2  = \frac{mh}{\ell_s^2} \star_{\rm CY} \left[ \text{Im}_{\rm CY}\wedge \delta_{O6} \right] = \frac{mh}{\ell_s^2}  \frac{{\cal V}_{\Pi_{O6}}}{{\cal V}_{\rm CY}}\, ,
\end{equation}
where in the last step we approximated the bump delta functions trough their constant Fourier modes, according to (\ref{deltasm}). Under the same assumptions, the only non-trivial flux EOM is the one of $H$ 
\begin{equation}\label{const2}
	d\star_{\rm CY}H = - 2 G_0^2 J_{\rm CY}^2 \, B  \left(C + \frac{1}{4}\right) = 0\,.
\end{equation}
Finally, the EOM of the dilaton and the Einstein equations take the form
\begin{subequations} 
	\begin{align}
		0 = & \; 12 \mu^2  \frac{e^{-2\phi}}{e^{2A}}-\frac{1}{2}e^{-2\phi}|H|^2+\frac{1}{4}|{G}_0|^2 - \frac{1}{4}|{G}_2|^2 - \frac{3}{4}|\hat{G}_4|^2 - \frac{5}{4}|\hat{G}_6|^2+ \frac{1}{4}e^{-\phi} \sum_i \delta^{(3)}_i \,, \label{EOMeindil1}\\
		\begin{split}
			0 = & \; \frac{1}{2}e^{-2\phi}\left(|H|^2_{mn} - \frac{1}{4}g_{mn}|H|^2\right) +\frac{1}{2}\left(|{G}_2|^2_{mn} - \frac{1}{8}g_{mn}|{G}_2|^2\right) \\
			& +\frac{1}{2}\left(|\hat{G}_4|^2_{mn} - \frac{3}{8}g_{mn}|\hat{G}_4|^2\right) +\frac{1}{2}\left(|\hat{G}_6|^2_{mn} - \frac{5}{8}g_{mn}|\hat{G}_6|^2\right) \\
			& +\frac{1}{16}g_{mn}|{G}_0|^2 + \frac{1}{2}e^{-\phi} \sum \left(\Pi_{i,mn}-\frac{7}{8}g_{mn}\right) \delta^{(3)}_i \,, \label{EOMeindil2}
		\end{split} \\
		0 = & -24 \mu^2 \frac{e^{-\phi}}{e^{2A}} - e^{-\phi}|H|^2 + \sum \delta^{(3)}_i \label{EOMeindil3} \,,
	\end{align}
\end{subequations}
where we have introduced $\mu^2 = 1/R^2$, as in the main text. Here one should make the replacement $e^\phi \to \left<e^\phi\right> = g_s$ and $e^A \to  \left<e^{A}\right> = 1$, and smear the delta functions as follows 
\begin{equation}
	\qquad \sum_i  \delta^{(3)}_i = \frac{mh}{\ell_s^2} \star_{\rm CY} \left[ \text{Im}_{\rm CY}\wedge \delta_{O6} \right] = \frac{mh}{\ell_s^2}  \frac{{\cal V}_{\Pi_{O6}}}{{\cal V}_{\rm CY}}\,.
\end{equation}
It is easy to verify that the smearing Ansatz (\ref{intfluxsm}) implies in our conventions 
\begin{equation}
	\begin{split} \label{fieldstr}
		& |H|^2 = 144 A^2 g_s^2 {G}_0^2 \,, \quad |{G}_2|^2 = 3 B^2 {G}_0^2 \,, \quad |\hat{G}_4|^2 = 12 C^2 {G}_0^2 \,, \quad |\hat{G}_6|^2 = 0 \,, \\
		&  |H|^2_{mn} = \frac{1}{2}g_{mn}|H|^2 \,, \quad |{G}_2|^2_{mn} = \frac{1}{3}g_{mn}|{G}_2|^2  \,, \quad |\hat{G}_4|^2_{mn} = \frac{2}{3}g_{mn}|\hat{G}_4|^2  \,, \quad |\hat{G}_6|^2_{mn} = 0 \,.
	\end{split}
\end{equation}
Combining (\ref{EOMeindil1}) and (\ref{EOMeindil3}) and exploiting (\ref{fieldstr}) we obtain
\begin{subequations} \label{App10dEOMeq01}
	\begin{align}
		\mu^2 & = \frac{{G}_0^2g_s^2}{72} \left( 144A^2 +3B^2 +36C^2-1\right)\, , \\
		\frac{mh}{\ell_s^{2}}\frac{{\cal V}_{\Pi_{\rm O6}}}{{\cal V}_{\rm CY}} & = \frac{{G}_0^2g_s}{3} \left( 576A^2 +3B^2 +36C^2-1\right)\, \label{const3} .
	\end{align}
\end{subequations}
Replacing (\ref{App10dEOMeq01}) into (\ref{EOMeindil2}) an taking the trace we find  
\begin{equation}
	\frac{mh}{\ell_s^{2}}\frac{{\cal V}_{\Pi_{\rm O6}}}{{\cal V}_{\rm CY}} = \frac{{G}_0^2g_s}{6} \left( 1584A^2 +3B^2 +84C^2-5\right)\,. \label{const4}
\end{equation}
Summarising, equations \ref{const1}, \ref{const2}, \ref{const3} and \ref{const4} tell us that a vacua must satisfy 
\begin{align}
	& 24 A = \frac{1}{3} \left( 576A^2 +3B^2 +36C^2-1\right) = \frac{1}{6} \left( 1584A^2 +3B^2 +84C^2-5\right) \,, \\
	& B\left(C+\frac{1}{4}\right) = 0 \,.
\end{align} 

\subsection{First-order corrections}

Away from the smearing approximation the dilaton and the warping factor are no more constant. Inspired by the results of \cite{Marchesano:2020qvg}, we assume the following Ansatz for the warp factor and the dilaton
\begin{equation}\label{ansatzDW}
	e^{-A}  = 1 + g_s \varphi + \cO(g_s^2) \, , \qquad e^{\phi}   = g_s \left(1 - 3  g_s \varphi\right) + \cO(g_s^3)\, ,
\end{equation} 
with $\varphi$ a real function. The metric $g_{mn}$ is no more Calabi--Yau, but we assume that its departure from the Calabi--Yau conditions can be described by a series of $\cO(g_s^n)$ corrections. At the level of the first correction there still exist a three-form $\Omega$ and a two-form $J$ such that 
\begin{equation}
	\star_6\, \text{Re}(\Omega) = \text{Im}(\Omega) + \mathcal{O}(g_s^2) \label{Tromega}\,, \qquad \star_6 \, J = -\frac{1}{2}\, J^2 + \mathcal{O}(g_s^2)\,,
\end{equation}
with $\star_6$ the corrected Hodge star operator. Inspired again by \cite{Marchesano:2020qvg} we assume moreover that $\Omega$ and $J$ satisfy
\begin{subequations} 
	\begin{align}
		& \text{Re}(\Omega) = \text{Re}(\Omega_{\rm CY})(1-g_s \varphi) + g_s K + \mathcal{O}(g_s^2) \,, \label{Reomega}\\
		& \text{Im}(\Omega) = \text{Im}(\Omega_{\rm CY}) (1+g_s\varphi) - g_s \star_{\rm CY} K + \mathcal{O}(g_s^2) \,,  \label{Imomega}\\
		& J = J_{\rm CY} + \mathcal{O}(g_s^2)\,, \label{eq017}
	\end{align}
\end{subequations}
with $K$ a current three-form such that 
\begin{equation}\label{kdef}
	\Delta_{\rm CY} K =  6 A g_s {G}_0^2\text{Re}(\Omega_{\rm CY})- \frac{mh}{\ell_s^2} \delta_{O6} + \mathcal{O}(g_s^2)\,,
\end{equation}
where $\Delta_{\rm CY}$ is the Laplace operator associated to the uncorrected Calabi--Yau metric (notice that such current can be easily build, see the construction in \cite{Hitchin:1999fh}). 

We will show now explicitly that the fluxes (\ref{solutionflux}) satisfy the EOMs and the Bianchi identities for certain values of the real constants $S$ and $R$ therein. We start discussing the Bianchi identities. According to Hodge theory, a form admits a unique decomposition in exact, harmonic and co-exact components.\footnote{Notice that such a decomposition depends on the metric. We define it with respect to the Calabi--Yau one.} The Ansatz (\ref{solutionflux}) tell us that neither $\hat{G}_6$ nor $H$ have a co-exact part and takes $G_0$ as a constant, therefore (\ref{BiG0}), (\ref{BiG6}) and (\ref{BiH}) are automatically satisfied. According to \cite{Marchesano:2020qvg}, the most general $K$ that satisfies (\ref{kdef}) is a closed three-form current which can be written as $K = \tilde{\varphi} \text{Re}(\Omega_{\rm CY})+c \,\text{Im}(\Omega_{\rm CY})+ \text{Re}(k)$ with $k$ a (2,1) primitive current. The requirement of $K$ being closed implies that $k_{2,1}$ must satisfy  (see \cite{Casas:2022mnz} for more details on the constraints that $\tilde{\varphi}$ and $k$ must satisfy)
\begin{subequations}\label{kconditions}
	\begin{equation} 
		k_{2,1} = \partial k_{1,1} + \bar{\partial} k_{2,0} + k_{2,1}^{h}\,,\qquad
		\partial k_{2,1} = \bar{\partial} \tilde{\varphi}\, \Omega_{\rm CY} \,, \qquad
		\text{Re}(\bar{\partial}k_{2,1}) = 0 \,.
	\end{equation}
\end{subequations}    
If we set $c=0$ and we require that $\varphi = \tilde{\varphi}$ it is straightforward to obtain
\begin{equation} \label{coDK1}
	d^\dagger_{\rm CY} K = \star_{\rm CY} d \left[2\varphi \text{Im}(\Omega_{\rm CY})\right] - V_{1,1}\,,
\end{equation} 
with $V_{1,1} = \star_{\rm CY} \left( \partial \bar{\partial} k_{1,1}\right)$ a primitive (1,1)-form. Exploiting the K\"ahler identity $\left[d^c, J \cdot \right] = d^\dagger$ with $d^c = -i(\partial - \bar{\partial})$ it is possible to remove $V_{1,1}$ from (\ref{coDK1}) and we obtain
\begin{equation}
	d^\dagger_{\rm CY} K = -J\cdot d \left[4\varphi \text{Im}(\Omega_{\rm CY})-\star_{\rm CY}K\right]\,,
\end{equation}
which implies that the Bianchi identity (\ref{BiG2}) is satisfied at first order in $g_s'$, we have indeed
\begin{equation}
	d{G}_2 = dd^\dagger_{\rm CY} K = \Delta_{\rm CY}K = {G}_0\wedge H - 4 \d_{\rm O6} +   N_\a \d_{\rm D6}^\a + \mathcal{O}(g_s^2)\,.
\end{equation}
Differentiating $\hat{G}_4$ we obtain instead
\begin{equation}
	\begin{split}
		d\hat{G}_4  & = 12 A g_s {G}_0 \left\{\star_{\rm CY} \left[d\varphi \wedge \text{Im}(\Omega_{\rm CY})\right]\wedge\text{Re}(\Omega_{\rm CY})\right\} \\ & = d^\dagger_{\rm CY}K \wedge H + \mathcal{O}(g_s^2) = {G}_2 \wedge H + \mathcal{O}(g_s^2)\, .
	\end{split}
\end{equation} 
In the first step we used the identity (see \cite[prop. 1.2.31]{huybrechts2005} adapted to the conventions of \cite{Tomasiello:2022dwe})
\begin{equation} \label{IdentityP}
	\star_{\rm CY} \left[J^s \wedge \alpha \right] = (-1)^{\frac{k(k+3)}{2}+1}\, \frac{s!}{(3-k-s)!}J^{3-k-s} \wedge I\left( \alpha\right)\,,
\end{equation}
where $\alpha$ is a primitive $k$-form and $I$ is the operator 
\begin{equation}
	I = \sum_{p,q = 0}^3 i^{p-q} \, \Pi^{p,q} \,,
\end{equation}  
to obtain $J^2_{\rm CY} \wedge d\varphi = -2 \star_{\rm CY} d^c \varphi$. In the second step we used equation (\ref{coDK1}). 

Let us now discuss the fluxes EOMs. The EOM of $\hat{G}_6$ (\ref{eomG6}) is trivially satisfied. Exploiting that $\hat{G}_6 = 0$ the EOM of $\hat{G}_4$ (\ref{eomG4}) becomes
\begin{equation}
	d \left[e^{4A} \star_{6} \hat{G}_4 \right] = 0\,.
\end{equation}  
Exploiting the fact that the exact and co-exact part of $G_4$ are of order $\mathcal{O}(g_s)$ and that $g_s \star_6 = g_s \star_{\rm CY} + \mathcal{O}(g_s^2)$ it further reduces to 
\begin{equation} \label{G4eq1}
	4\left(2C+6A\right) g_s  {G}_0 \, d \varphi \wedge J_{\rm CY} + d \star_{\rm CY} d \left[ S g_s^{-1} J_{\rm CY} \wedge \text{Im}(v)\right] = 0 + \mathcal{O}(g_s^2)\,.
\end{equation} 
Using the identity (\ref{IdentityP}) it is straightforward to prove that $ \star_{\rm CY} \left( J_{\rm CY} \wedge \text{Im}(v) \right) $ is a closed form. Exploiting the definition of $v$ and $f_*$ (\ref{ansatzVF}) the last term of (\ref{G4eq1}) takes the form
\begin{equation}
	d \star_{\rm CY} d \left[ \frac{S}{g_s}  J_{\rm CY} \wedge \text{Im}(v)\right] = \frac{S}{2} \star_{\rm CY}^{-1}\left(J_{\rm CY} \wedge d_c \Delta_{\rm CY} f_* \right) = -4 S g_s {G}_0 \, d \varphi \wedge J_{\rm CY} \,.
\end{equation}
The EOM of $\hat{G}_4$ is then simply
\begin{equation}\label{cond2}
	4\left(2C+6A-S\right) g_s  {G}_0 \, d \varphi \wedge J_{\rm CY} = 0 + \mathcal{O}(g_s^2)\,.
\end{equation}
The EOM of ${G}_2$ (\ref{eomG2}) reduces to 
\begin{equation}
	d \star_{\rm CY} {G}_2 = 0 + \mathcal{O}(g_s)\, ,
\end{equation}
which is trivially satisfied because ${G}_2$ does not have an exact part. The EOM of $H$ becomes
\begin{equation}\label{eomHstep}
	d(g_s^{-2}(1+2g_s \varphi)\star_6 H) + \star_{\rm CY}\hat{G}_4 \wedge{G}_2 + \star_{\rm CY} {G}_2 \wedge {G}_0 = 0 + \mathcal{O}(g_s) \,, 
\end{equation}
which we can evaluate term by term. Combining the ansatz (\ref{Reomega}) and (\ref{Imomega}) with the transformation property (\ref{Tromega}) we obtain
\begin{equation} 
	\star_6 \text{Re}(\Omega_{\rm CY}) = \text{Im}(\Omega_{\rm CY})(1+2\varphi g_s) - 2 g_s \star_{\rm CY} K + \mathcal{O}(g_s^2)\,,
\end{equation}
and the first term of (\ref{eomHstep}) becomes
\begin{equation}
	\begin{split}
	(1) \, = &\;  24 \, A \, {G}_0 d\varphi \wedge \text{Im}(\Omega_{\rm CY}) - 6 (2A - AR) {G}_0 \star_{\rm CY} d^\dagger_{\rm CY} K \\ & -\frac{S}{2g_s^2} d\star_{\rm CY}   d\re \left(\bar{v} \cdot \Omega_{\rm CY} \right) + \mathcal{O}(g_s)\,,
	\end{split}
\end{equation}
where there latter term further reduce to
\begin{equation}
	 -\frac{S}{2g_s^2} d\star_{\rm CY}   d\re \left(\bar{v} \cdot \Omega_{\rm CY} \right) = - 4 {G}_0 S d\varphi \wedge \text{Im}(\Omega_{\rm CY}) \,.
\end{equation}
The second term of (\ref{eomHstep}) becomes
\begin{equation}
	(2) = - 2 \, C \, {G}_0 \left(B\, {G}_0 J_{\rm CY}^2  - 4 d\varphi \wedge \text{Im}(\Omega_{\rm CY}) + \star_{\rm CY} d^\dagger_{\rm CY}K \right) + \mathcal{O}(g_s) \,,
\end{equation}
where we used the K\"ahler identity $\left[J\cdot, J\wedge\right] = H$ with $H$ the operator that on $k$-forms act as $H \alpha = (3-k)\alpha$. The third term of (\ref{eomHstep}) becomes simply
\begin{equation} 
		(3) = {G}_0 \star_{\rm CY} d^\dagger_{\rm CY} K_3 - \frac{1}{2} B {G}_0 J_{\rm CY}^2 + \mathcal{O}(g_s) \,.
\end{equation}
Replacing everything in the EOM of $H$, equation (\ref{eomH}) finally becomes
\begin{equation}\label{cond1}
		\begin{split}
			& \left(24 A + 8 C - 4 S\right) {G}_0 \, d\varphi \wedge \text{Im}(\Omega_{\rm CY}) + {G}_0 (1-2C - 12 A + 6A R) \star_{\rm CY} d^\dagger_{\rm CY} K \\  & - 2 G_0^2 \, B \left( C + \frac{1}{4}\right) = 0 + \mathcal{O}(g_s) \,.
		\end{split}
\end{equation}  

Summarising, equations (\ref{cond2}) and (\ref{cond1}) tell us that a vacua of the form (\ref{solutionflux}) exist provided that $S$ and $R$ satisfy
\begin{equation}
	S = 2C+6A\,, \qquad 6AR = 12A + 2C - 1\,. 
\end{equation} 


\section{DBI computation}
\label{ap:DBI}

In this appendix we derive the expressions for the DBI action that are used in section \ref{s:membranes} to compute the tension of D8-branes and D6-branes with internal worldvolume fluxes. 

Let us start by considering a D8-brane wrapping the whole of a Calabi--Yau manifold $X_6$. Ignoring curvature corrections, the contribution to the DBI coming from the internal dimensions involves the square root of 
\begin{equation}
\label{eq:dbii}
\operatorname{det}\left(g_{a b}-\mathcal{F}_{a b}\right)=\operatorname{det} g \operatorname{det}(\mathbb{I}+A)=\operatorname{det} g\left(1-\frac{t_{2}}{2}+\frac{t_{2}^{2}}{8}-\frac{t_{4}}{4}+\operatorname{det} A\right)\, ,
\end{equation}
where we have used the Cayley–Hamilton theorem and introduced the definitions $A\equiv -g^{-1}\mathcal{F}$ and $t_k\equiv \text{Tr}\left(A^k\right)$. Assuming that $\cF$ is a (1,1)-form it follows that
\begin{align}
-\frac{1}{2} t_{2} &=\left(\frac{1}{2} \mathcal{F} \wedge J \wedge J\right)^{2} +\left(\mathcal{F} \wedge \mathcal{F} \wedge J\right) \cdot d\text{vol}_{X_6}\, , \\
\left(\frac{t_{2}^{2}}{8}-\frac{t_{4}}{4}\right) &=\left[\left(\frac{1}{2} \mathcal{F} \wedge \mathcal{F} \wedge J\right)^{2} -2\left(\frac{1}{2} \mathcal{F} \wedge J \wedge J\right) \cdot\left(\frac{1}{6} \mathcal{F} \wedge \mathcal{F} \wedge \mathcal{F}\right)\right] \, ,\\
\operatorname{det} A &=\frac{1}{36}\left(\mathcal{F}\wedge\mathcal{F}\wedge\mathcal{F}\right)^2 \, ,
\end{align} 
where $d\text{vol}_{X_6}=-\frac{1}{6}J\wedge J\wedge J$ and the product means contraction of two six-form with the metric. Putting everything together,  the integrand of the  DBI action  for a D8 wrapping the whole of $X_6$  and with a $\left(1,1\right)$ worldvolume flux on it  can be written as
\begin{equation}
d{\rm DBI}_{\rm D8} = g_s^{-1}
\sqrt{\left(\frac{1}{6} J \wedge J \wedge J-\frac{1}{2} \mathcal{F} \wedge \mathcal{F} \wedge J\right)^{2}+\left(\frac{1}{6} \mathcal{F} \wedge \mathcal{F} \wedge \mathcal{F}-\frac{1}{2} \mathcal{F} \wedge J \wedge J\right)^{2}} d\text{vol}_{X_6}\, .
\label{ap:DBID8}
\end{equation}	
Therefore, whenever $\mathcal{F}\wedge\mathcal{F}=3 J\wedge J$ we obtain a perfect square, signalling that we have a BPS configuration. This is just a  particular solution of the MMMS equations \cite{Marino:1999af}, which in our conventions read
\begin{align}
\frac{1}{6} \mathcal{F} \wedge \mathcal{F} \wedge \mathcal{F}-\frac{1}{2} \mathcal{F} \wedge J \wedge J
=\tan \theta\,  \left(\frac{1}{6} J \wedge J \wedge J - \frac{1}{2} \mathcal{F} \wedge \mathcal{F} \wedge J\right)\, ,
\end{align}
with $\theta$ defined as in \eqref{calibration}.

We can apply the same reasoning considering to a D6-brane wrapping an internal 4-cycle $\mathcal{S}$ of $X_6$. In this case, the determinant that appears in the DBI action can be expressed as
\begin{equation}
\label{eq:dbii6}
\operatorname{det}\left(g_{a b}-\mathcal{F}_{a b}\right)=\operatorname{det} g \operatorname{det}(\mathbb{I}+A)=\operatorname{det} g\left(1-\frac{t_{2}}{2}+\operatorname{det} A\right)\, ,
\end{equation}
Assuming that $\cF$ is a (1,1)-form and denoting by $J$ the pull-back of $J_{\rm CY}$ on ${\cal S}$ we have that 
\begin{align}
-\frac{1}{2} t_{2} &=\left( \mathcal{F} \wedge J\right)^{2}  +\left(\mathcal{F} \wedge \mathcal{F}\right) \cdot d\text{vol}_{\mathcal{S}}\, , \\
\operatorname{det} A &=\left(\frac{1}{2}\mathcal{F}\wedge\mathcal{F}\right)^2 \, ,
\end{align}
where $d\text{vol}_{\mathcal{S}}=-\frac{1}{2}J\wedge J$. Taking into account all these terms,  the internal part of the  DBI action  for a D6 wrapping a four-cycle $\mathcal{S}$ and with  a $\left(1,1\right)$ internal worldvolume flux on it reads
\begin{align}
d{\rm DBI}_{\rm D6} = g_s^{-1} \sqrt{\left(-\frac{1}{2} J \wedge J+\frac{1}{2} \mathcal{F} \wedge \mathcal{F}\right)^{2}+(J \wedge \mathcal{F})^{2}}\;d\text{vol}_{\mathcal{S}}\, .
\label{ap:DBID6}
\end{align}
We see that for $J\wedge J = \cF \wedge \cF$ the interior of the square root becomes a perfect square. Accordingly, we recover again a solution of the MMMS equations \cite{Marino:1999af}, which for the case at hand read 
\begin{align}
\tan^{-1}\theta  \left(J \wedge \mathcal{F}\right)= \frac{1}{2} J \wedge J-\frac{1}{2} \mathcal{F} \wedge \mathcal{F}\, .
\end{align}




\bibliographystyle{JHEP2015}
\bibliography{papers}

\end{document}